
\newcount\mgnf\newcount\tipi\newcount\tipoformule
\newcount\aux\newcount\driver\newcount\cind\global\newcount\bz
\newcount\tipobib\newcount\stile\newcount\modif
\newcount\noteno\noteno=1\newcount\pos\newcount\figure

\newdimen\stdindent\newdimen\bibskip
\newdimen\maxit\maxit=0pt

\stile=0         
\tipobib=1       
\bz=0            
\cind=0          
\mgnf=1          
\tipoformule=0   
\aux=1           
\pos=1		 
\figure=1        


\ifnum\mgnf=0
   \magnification=\magstep0 
   \hsize=17truecm\vsize=24truecm\hoffset=-0.5truecm\voffset=-1.0truecm
   \parindent=4.pt\stdindent=\parindent\fi
\ifnum\mgnf=1
   \magnification=\magstep1\hoffset=-0.5truecm
   \voffset=-1.0truecm\hsize=18truecm\vsize=24.truecm
   \baselineskip=14pt plus0.1pt minus0.1pt \parindent=6pt
   \lineskip=4pt\lineskiplimit=0.1pt      \parskip=0.1pt plus1pt
   \stdindent=\parindent\fi
\ifnum\mgnf=2
   \magnification=\magstep2\hoffset=-1.5truecm
   \voffset=-1.0truecm\hsize=19truecm\vsize=24.truecm
   \baselineskip=14pt plus0.1pt minus0.1pt \parindent=6pt
   \lineskip=4pt\lineskiplimit=0.1pt\fi


\def\fine#1{}
\def\draft#1{\bz=1\ifnum\mgnf=1\baselineskip=22pt 
   \else\baselineskip=16pt\fi
   \ifnum\stile=0\headline={\hfill DRAFT #1}\fi\raggedbottom
    \setbox150\vbox{\parindent=0pt\centerline{\bf Figures' captions}\*}
    \def\gnuins ##1 ##2 ##3{\gnuinsf {##1} {##2} {##3}}
    \def\gnuin ##1 ##2 ##3 ##4 ##5 ##6{\gnuinf {##1} {##2} {##3} 
		{##4} {##5} {##6}} 
    \def\eqfig##1##2##3##4##5##6{\eqfigf {##1} {##2} {##3} {##4} {##5} {##6}}
    \def\eqfigbis##1##2##3##4##5##6##7	
             {\eqfigbisf {##1} {##2} {##3} {##4} {##5} {##6} {##7}}
    \def\eqfigfor##1##2##3##4##5##6##7
             {\eqfigforf {##1} {##2} {##3} {##4} {##5} {##6} {##7}}
      \def\fine ##1{\vfill\eject\parindent=0pt
	 	\def\geq(####1){}
               \unvbox150\vfill\eject\raggedbottom
                \centerline{FIGURES}\unvbox149 ##1}}

\def\large{\draft{}\bz=0\headline={\hfill}}


\newcount\prau

\def\titolo#1{\setbox197\vbox{ 
\leftskip=0pt plus16em \rightskip=\leftskip
\spaceskip=.3333em \xspaceskip=.5em \parfillskip=0pt
\pretolerance=9999  \tolerance=9999
\hyphenpenalty=9999 \exhyphenpenalty=9999
\ftitolo #1}}
\def\abstract#1{\setbox198\vbox{
    \centerline{\vbox{\advance\hsize by -2cm \parindent=0pt\it Abstract: #1}}}}
\def\parole#1{\setbox195\hbox{
     \centerline{\vbox{\advance\hsize by -2cm \parindent=0pt Keywords: #1.}}}}
\def\autore#1#2{\setbox199\hbox{\unhbox199\ifnum\prau=0 #1%
\else, #1\fi\global\advance\prau by 1$^{\simbau}$}
     \setbox196\vbox{\advance\hsize by -\parindent\copy196\ottopunti\item{$^{\simbau}$}{#2}}}
\def\prima{\unvbox197\vskip1truecm\centerline{\unhbox199}\footnote{}{\unvbox196}
\vskip1truecm\unvbox198\vskip1truecm\copy195}

\def\simbau{\ifcase\prau
	\or \dagger \or \ddagger \or * \or \star \or \bullet\fi}


     \let\k=\kappa

\let\ge=\geq
\let\le=\leq


{\count255=\time\divide\count255 by 60 \xdef\oramin{\number\count255}
        \multiply\count255 by-60\advance\count255 by\time
   \xdef\oramin{\oramin:\ifnum\count255<10 0\fi\the\count255}}
\def\ora{\oramin }

\def\data{\number\day/\ifcase\month\or gennaio \or febbraio \or marzo \or
aprile \or maggio \or giugno \or luglio \or agosto \or settembre
\or ottobre \or novembre \or dicembre \fi/\number\year;\ \ora}

\setbox200\hbox{$\scriptscriptstyle \data $}


\newcount\pgn \pgn=1
\newcount\firstpage

\def\foglio{\number\numsec:\number\pgn\global\advance\pgn by 1}
\def\foglioa{A\number\numsec:\number\pgn\global\advance\pgn by 1}

\def\pagina{\vfill\eject}
\def\ppagina{\ifodd\pageno\pagina\null\pagina\else\pagina\fi}
\def\ppaginan{\ifodd-\pageno\pagina\null\pagina\else\pagina\fi}

\def\setind{\firstpage=\pageno}
\def\setcap#1{\null\def\titlecap{#1}\global\firstpage=\pageno}
\def\titletesi{Indici critici per sistemi fermionici in una dimensione}

\ifnum\stile=1
  \def\pagenumbers{\headline={%
  \ifnum\pageno=\firstpage\hfil\else%
     \ifodd\pageno\hfill{\sc\titlecap}~~{\bf\folio}%
      \else{\bf\folio}~~{\sc\titletesi}\hfill\fi\fi}
  \footline={\ifnum\bz=0
                   \hfill\else\rlap{\hbox{\copy200}\ $\st[\foglio]$}\hfill\fi}}
  \def\pagenumbersind{\headline={%
  \ifnum\pageno=\firstpage\hfil\else%
    \ifodd\pageno\hfill{\rm\romannumeral-\pageno}%
     \else{\rm\romannumeral-\pageno}\hfill\fi\fi}
  \footline={\ifnum\bz=0
                   \hfill\else\rlap{\hbox{\copy200}\ $\st[\foglio]$}\hfill\fi}}
\else
  \def\pagenumbers{\headline={\hfill}
     \footline={\ifnum\bz=0\hfill\folio\hfill
                \else\rlap{\hbox{\copy200}\ $\st[\foglio]$}
		   \hfill\folio\hfill\fi}}
\fi

\pagenumbers

\def\numeropag#1{
   \ifnum #1<0 \romannumeral -#1\else \number #1\fi
   }


\global\newcount\numsec\global\newcount\numfor
\global\newcount\numfig\global\newcount\numpar
\global\newcount\numteo\global\newcount\numlem

\numfig=1\numsec=0

\gdef\profonditastruttura{\dp\strutbox}
\def\senondefinito#1{\expandafter\ifx\csname #1\endcsname\relax}
\def\SIA #1,#2,#3 {\senondefinito{#1#2}%
\expandafter\xdef\csname#1#2\endcsname{#3}\else%
\write16{???? ma #1,#2 e' gia' stato definito !!!!}\fi}
\def\etichetta(#1){(\veroparagrafo.\veraformula)
\SIA e,#1,(\veroparagrafo.\veraformula)
 \global\advance\numfor by 1
\write15{\string\FU (#1){\equ(#1)}}
\9{ \write16{ EQ \equ(#1) == #1  }}}
\def \FU(#1)#2{\SIA fu,#1,#2 }
\def\etichettaa(#1){(A\veroparagrafo.\veraformula)
 \SIA e,#1,(A\veroparagrafo.\veraformula)
 \global\advance\numfor by 1
\write15{\string\FU (#1){\equ(#1)}}
\9{ \write16{ EQ \equ(#1) == #1  }}}
\def \FU(#1)#2{\SIA fu,#1,#2 }
\def\tetichetta(#1){\veroparagrafo.\veroteorema
\SIA e,#1,{\veroparagrafo.\veroteorema}
\global\advance\numteo by1
\write15{\string\FU (#1){\equ(#1)}}%
\9{\write16{ EQ \equ(#1) == #1}}}
\def\tetichettaa(#1){A\veroparagrafo.\veroteorema
\SIA e,#1,{A\veroparagrafo.\veroteorema}
\global\advance\numteo by1
\write15{\string\FU (#1){\equ(#1)}}%
\9{\write16{ EQ \equ(#1) == #1}}}
\def\letichetta(#1){\veroparagrafo.\verolemma
\SIA e,#1,{\veroparagrafo.\verolemma}
\global\advance\numlem by1
\write15{\string\FU (#1){\equ(#1)}}%
\9{\write16{ EQ \equ(#1) == #1}}}
\def\getichetta(#1){{\bf Fig. \verafigura}:
 \SIA e,#1,{\verafigura}
 \global\advance\numfig by 1
\write15{\string\FU (#1){\equ(#1)}}
\9{ \write16{ Fig. \equ(#1) ha simbolo  #1  }}}

\def\veroparagrafo{\number\numsec}\def\veraformula{\number\numfor}
\def\verafigura{\number\numfig}\def\veroteorema{\number\numteo}
\def\verolemma{\number\numlem}

\def\geq(#1){\getichetta(#1)\galato(#1)}
\def\Eq(#1){\eqno{\etichetta(#1)\alato(#1)}}
\def\eq(#1){&\etichetta(#1)\alato(#1)}
\def\Eqa(#1){\eqno{\etichettaa(#1)\alato(#1)}}
\def\eqa(#1){&\etichettaa(#1)\alato(#1)}
\def\teq(#1){\tetichetta(#1)\talato(#1)}
\def\teqa(#1){\tetichettaa(#1)\talato(#1)}
\def\leq(#1){\letichetta(#1)\talato(#1)}

\def\Eqr{\eqno(\veroparagrafo.\veraformula)\advance\numfor by 1}
\def\eqr{&(\veroparagrafo.\veraformula)\advance\numfor by 1}
\def\Eqar{\eqno(A\veroparagrafo.\veraformula)\advance\numfor by 1}
\def\eqar{&(A\veroparagrafo.\veraformula)\advance\numfor by 1}

\def\eqv(#1){\senondefinito{fu#1}$\clubsuit#1$\write16{Manca #1 !}%
\else\csname fu#1\endcsname\fi}
\def\equ(#1){\senondefinito{e#1}\eqv(#1)\else\csname e#1\endcsname\fi}


\newdimen\gwidth

\def\commenta#1{\ifnum\bz=1\strut \vadjust{\kern-\profonditastruttura
 \vtop to \profonditastruttura{\baselineskip
 \profonditastruttura\vss
 \rlap{\kern\hsize\kern0.1truecm
  \vbox{\hsize=1.7truecm\raggedright\nota\noindent #1}}}}\fi}
\def\talato(#1){\ifnum\bz=1\strut \vadjust{\kern-\profonditastruttura
 \vtop to \profonditastruttura{\baselineskip
 \profonditastruttura\vss
 \rlap{\kern-1.2truecm{$\scriptstyle#1$}}}}\fi}
\def\alato(#1){\ifnum\bz=1
 {\vtop to \profonditastruttura{\baselineskip
 \profonditastruttura\vss
 \rlap{\kern-\hsize\kern-1.2truecm{$\scriptstyle #1$}}}}\fi}
\def\galato(#1){\ifnum\bz=1 \gwidth=0pt 
 {\vtop to \profonditastruttura{\baselineskip
 \profonditastruttura\vss
 \rlap{\kern-\gwidth\kern-2.2truecm{$\scriptstyle#1$}}}}\fi}


\def\magg{\ifnum\mgnf=0\magstep0\else\magstep1\fi}

\newskip\ttglue

\font\ftitolo=cmbx12 
\font\eighttt=cmtt8 \font\sevenit=cmti7  \font\sevensl=cmsl8
\font\sc=cmcsc10

\font\msytwww=msbm7 scaled\magstep1


\def\settepunti{\def\rm{\fam0\sevenrm}
\textfont0=\sevenrm \scriptfont0=\fiverm \scriptscriptfont0=\fiverm
\textfont1=\seveni \scriptfont1=\fivei   \scriptscriptfont1=\fivei
\textfont2=\sevensy \scriptfont2=\fivesy   \scriptscriptfont2=\fivesy
\textfont3=\tenex \scriptfont3=\tenex   \scriptscriptfont3=\tenex
\textfont\itfam=\sevenit  \def\it{\fam\itfam\sevenit}%
\textfont\slfam=\sevensl  \def\sl{\fam\slfam\sevensl}%
\textfont\ttfam=\eighttt  \def\tt{\fam\ttfam\eighttt}
\textfont\bffam=\sevenbf 
\scriptfont\bffam=\fivebf \scriptscriptfont\bffam=\fivebf  
\def\bf{\fam\bffam\sevenbf}%
\tt \ttglue=.5em plus.25em minus.15em
\setbox\strutbox=\hbox{\vrule height6.5pt depth1.5pt width0pt}%
\let\sc=\fiverm \normalbaselines\rm
\ifnum\mgnf=0\baselineskip=8pt\normalbaselineskip=8pt
\else\baselineskip=14pt\normalbaselineskip=14pt\fi}

\font\eightrm=cmr8 scaled \magg
\font\eightbf=cmbx8 scaled \magg
\font\eightit=cmti8 scaled \magg
\font\eightsl=cmsl8 scaled \magg
\font\eighttt=cmtt8 scaled \magg
\font\eightsy=cmsy8 scaled \magg
\font\eighti=cmmi8 scaled \magg
\font\sixrm=cmr6 scaled \magg
\font\sixbf=cmbx6 scaled \magg

\font\sixsy=cmsy6 scaled \magg
\font\sixi=cmmi6 scaled \magg

\def\ottopunti{\def\rm{\fam0\eightrm}%
\textfont0=\eightrm\scriptfont0=\sixrm\scriptscriptfont0=\fiverm%
\textfont1=\eighti\scriptfont1=\sixi\scriptscriptfont1=\fivei%
\textfont2=\eightsy\scriptfont2=\sixsy\scriptscriptfont2=\fivesy%
\textfont3=\tenex\scriptfont3=\tenex\scriptscriptfont3=\tenex%
\textfont\itfam=\eightit\def\it{\fam\itfam\eightit}
\textfont\slfam=\eightsl\def\sl{\fam\slfam\eightsl}%
\textfont\ttfam=\eighttt\def\tt{\fam\ttfam\eighttt}%
\textfont\bffam=\eightbf %
\scriptfont\bffam=\sixbf\scriptscriptfont\bffam=\fivebf %
\def\bf{\fam\bffam\eightbf}%
\ttglue=.5em plus.25em minus.15em%
\setbox\strutbox=\hbox{\vrule height6.5pt depth1.5pt width0pt}%
\let\sc=\fiverm \normalbaselines\rm
\ifnum\mgnf=0\baselineskip=8pt\normalbaselineskip=8pt
\else\baselineskip=18pt\normalbaselineskip=18pt\fi}%

\let\nota=\ottopunti

\def\nnn{\hbox{\msytwww N}} 
 \def\zzz{\hbox{\msytwww Z}}


\font\tenmib=cmmib10
\font\sevenmib=cmmib10 scaled 800

\textfont5=\tenmib  \scriptfont5=\sevenmib  \scriptscriptfont5=\fivei

\mathchardef\aaa= "050B
\mathchardef\xxx= "0518
\mathchardef\oo = "0521
\mathchardef\Dp = "0540
\mathchardef\H  = "0548
\mathchardef\FFF= "0546
\mathchardef\ppp= "0570
\mathchardef\nnn= "0517

\newdimen\xshift \newdimen\xwidth \newdimen\yshift \newdimen\ywidth
\newdimen\laln

\ifnum\pos=0\def\midinsert{}\def\endinsert{}\fi

\def\ins#1#2#3{\nointerlineskip\vbox to0pt {\kern-#2 \hbox{\kern#1 #3}
\vss}}

\def\lineauno#1#2#3#4{
\xwidth=#1 \xshift=\hsize \advance\xshift 
by-\xwidth \divide\xshift by 2
\yshift=#2 \divide\yshift by 2
\parindent=0pt
\line{\hglue\xshift \vbox to #2{\hsize #1\vfil 
#3 \includegraphics{#40.ps}
}\hfill}}

\def\unafig#1#2#3#4#5#6{
\setbox99\vbox{\parindent=0pt\par\lineauno{#1}{#2}{#3}{#4}
\nobreak
\smallskip
\didascalia{\geq(#6)#5}}}

\def\eqfig#1#2#3#4#5#6{
\unafig{#1}{#2}{#3}{#4}{#5}{#6}
\midinsert\unvbox99\endinsert
}

\def\eqfigf#1#2#3#4#5#6{
\unafig{#1}{#2}{#3}{#4}{#5}{#6}
\midinsert\unvbox99\endinsert
\setbox149\vbox{\unvbox149 \*\* \centerline{Fig. \equ(#6)} 
\nobreak
\*
\nobreak
\lineauno{#1}{#2}{#3}{#4}}
\setbox150\vbox{\unvbox150 \parindent=0pt\*{\bf Fig. \equ(#6)}: #5}
}

\def\lineadue#1#2#3#4#5{
\xwidth=#1 \multiply\xwidth by 2 
\xshift=\hsize \advance\xshift 
by-\xwidth \divide\xshift by 3
\yshift=#2 \divide\yshift by 2
\ywidth=#2
\parindent=0pt
\line{\hfill
\vbox to \ywidth{\vfil #3 \includegraphics{#50.ps}}
\hskip\xshift%
\vbox to \ywidth{\vfil #4 \includegraphics{#51.ps}}}}

\def\duefig#1#2#3#4#5#6#7{
\setbox99\vbox{\parindent=0pt\par\lineadue{#1}{#2}{#3}{#4}{#5}
\nobreak
\*\*
\didascalia{\geq(#7)#6}}}

\def\eqfigbisf#1#2#3#4#5#6#7{
\duefig{#1}{#2}{#3}{#4}{#5}{#6}{#7}
\midinsert\unvbox99\endinsert
\setbox149\vbox{\unvbox149 \*\* \centerline{Fig. \equ(#7)} 
\nobreak
\*
\nobreak
\lineadue{#1}{#2}{#3}{#4}{#5}}
\setbox150\vbox{\unvbox150 \parindent=0pt\*{\bf Fig. \equ(#7)}: #6\*}
}

\def\eqfigbis#1#2#3#4#5#6#7{
\duefig{#1}{#2}{#3}{#4}{#5}{#6}{#7}
\midinsert\unvbox99\endinsert
}

\def\dimenfor#1#2{\par\xwidth=#1 \multiply\xwidth by 2 
\xshift=\hsize \advance\xshift 
by-\xwidth \divide\xshift by 3
\divide\xwidth by 2 
\yshift=#2 
\ywidth=#2}

\def\lineaquattro#1#2#3#4#5{
\parindent=0pt
\hbox to \hsize{\hskip\xshift 
\hbox to \xwidth{\vbox to \ywidth{\vfil#1\includegraphics{#50.ps}}\hfill}%
\hskip\xshift%
\hbox to \xwidth{\vbox to \ywidth{\vfil#2\includegraphics{#51.ps}}\hfill}\hfill}
\nobreak
\line{\hglue\xshift 
\hbox to \xwidth{\vbox to \ywidth{\vfil #3 \includegraphics{#52.ps}}\hfill}%
\hglue\xshift
\hbox to \xwidth{\vbox to\ywidth {\vfil #4 \includegraphics{#53.ps}}\hfill}\hfill}}

\def\quattrofig#1#2#3#4#5#6#7{
\setbox99\vbox{\parindent=0pt\par\lineaquattro{#1}{#2}{#3}{#4}{#5}
\nobreak
\*\*
\didascalia{\geq(#7)#6}}}

\def\eqfigforf#1#2#3#4#5#6#7{
\quattrofig{#1}{#2}{#3}{#4}{#5}{#6}{#7}
\midinsert\unvbox99\endinsert
\setbox149\vbox{\unvbox149 \*\* \centerline{Fig. \equ(#7)} 
\nobreak
\*
\nobreak
\lineadue{#1}{#2}{#3}{#4}{#5}}
\setbox150\vbox{\unvbox150 \parindent=0pt\*{\bf Fig. \equ(#7)}: #6\*}}

\def\eqfigfor#1#2#3#4#5#6#7{
\quattrofig{#1}{#2}{#3}{#4}{#5}{#6}{#7}
\midinsert\unvbox99\endinsert
}

\def\eqfigter#1#2#3#4#5#6#7{
\line{\hglue\xshift 
\vbox to \ywidth{\vfil #1 \includegraphics{#2.ps}}
\hglue30pt
\vbox to \ywidth{\vfil #3 \includegraphics{#4.ps}}\hfill}
\multiply\xshift by 3 \advance\xshift by \xwidth \divide\xshift by 2
\line{\hfill\hbox{#7}}
\line{\hglue\xshift 
\vbox to \ywidth{\vfil #5 \includegraphics{#6.ps}}}}


\def\7{\ifnum\modif=1\write13\else\write12\fi}
\def\8{\immediate\write13}


\def\gnuin #1 #2 #3 #4 #5 #6{\midinsert\vbox{\vbox to 260pt{
\hbox to 420pt{
\hbox to 200pt{\hfill\nota (a)\hfill}\hfill
\hbox to 200pt{\hfill\nota (b)\hfill}}
\vbox to 110pt{\vfill\hbox to 420pt{
\hbox to 200pt{\includegraphics{#1.ps}\hfill}\hfill
\hbox to 200pt{\includegraphics{#2.ps}\hfill}
}}\vfill
\hbox to 420pt{
\hbox to 200pt{\hfill\nota (c)\hfill}\hfill
\hbox to 200pt{\hfill\nota (d)\hfill}}
\vbox to 110pt{\vfill\hbox to 420pt{
\hbox to 200pt{\includegraphics{#3.ps}\hfill}\hfill
\hbox to 200pt{\includegraphics{#4.ps}\hfill}
}}\vfill}
\vskip0.25cm
\0\didascalia{\geq(#5): #6}}
\endinsert}

\def\gnuinf #1 #2 #3 #4 #5 #6{\midinsert\nointerlineskip\vbox to 260pt{
\hbox to 420pt{
\hbox to 200pt{\hfill\nota (a)\hfill}\hfill
\hbox to 200pt{\hfill\nota (b)\hfill}}
\vbox to 110pt{\vfill\hbox to 420pt{
\hbox to 200pt{\includegraphics{#1.ps}\hfill}\hfill
\hbox to 200pt{\includegraphics{#2.ps}\hfill}
}}\vfill
\hbox to 420pt{
\hbox to 200pt{\hfill\nota (c)\hfill}\hfill
\hbox to 200pt{\hfill\nota (d)\hfill}}
\vbox to 110pt{\vfill\hbox to 420pt{
\hbox to 200pt{\includegraphics{#3.ps}\hfill}\hfill
\hbox to 200pt{\includegraphics{#4.ps}\hfill}
}}\vfill}
\?
\0\didascalia{\geq(#5): #6}
\endinsert
\global\setbox150\vbox{\unvbox150 \*\*\0 Fig. \equ(#5): #6}
\global\setbox149\vbox{\unvbox149 \*\*
    \vbox{\centerline{Fig. \equ(#5)(a)} \nobreak
    \vbox to 200pt{\vfill\includegraphics{#1.ps_f}}}\*\*
    \vbox{\centerline{Fig. \equ(#5)(b)}\nobreak
    \vbox to 200pt{\vfill\includegraphics{#2.ps_f}}}\*\*
    \vbox{\centerline{Fig. \equ(#5)(c)}\nobreak
    \vbox to 200pt{\vfill\includegraphics{#3.ps_f}}}\*\*
    \vbox{\centerline{Fig. \equ(#5)(d)}\nobreak
    \vbox to 200pt{\vfill\includegraphics{#4.ps_f}}}
}}

\def\gnuins #1 #2 #3{\midinsert\nointerlineskip
\vbox{\line{\hfill\ifnum\figure=1 \vbox to 310pt{\vfill
\includegraphics{#1.ps}\vskip .4truecm}\hfill\fi\hglue425pt\hfill}
\0\didascalia{\geq(#2)#3}}
\endinsert}

\def\gnuinsf #1 #2 #3{\midinsert\nointerlineskip
\vbox{\line{\hfill\ifnum\figure=1\vbox to 310pt{\vfill
\includegraphics{#1.ps}\vskip .4truecm}\hfill\fi\hglue380pt\hfill}
\0\didascalia{\geq(#2)#3}}\endinsert
\global\setbox150\vbox{\unvbox150 \*\0{\bf Fig. \equ(#2)}: #3}
\global\setbox149\vbox{\unvbox149 \*\* 
    \vbox{\centerline{Fig. \equ(#2)}\nobreak
    \ifnum\figure=1\vbox to 300pt{\vfill\includegraphics{#1.ps}}\fi}} 
}


\def\9#1{\ifnum\aux=1#1\else\relax\fi}
\let\numero=\number
\def\boh{\hbox{$\clubsuit$}\write16{Qualcosa di indefinito a pag. \the\pageno}}
\def\didascalia#1{\vbox{\nota\ifnum\mgnf=0\baselineskip=10pt\normalbaselineskip=10pt
\else\baselineskip=16pt\normalbaselineskip=16pt\fi
\0#1\hfill}\vskip0.3truecm}
\def\frac#1#2{{#1\over #2}}
\def\V#1{\underline{#1}}
	
		\let\i=\infty
		
 	\let\0=\noindent
\def\guida{\leaders\hbox to 1em{\hss.\hss}\hfill}
\def\tende#1{\vtop{\ialign{##\crcr\rightarrowfill\crcr
              \noalign{\kern-1pt\nointerlineskip}
              \hglue3.pt${\scriptstyle #1}$\hglue3.pt\crcr}}}
\def\otto{{\kern-1.truept\leftarrow\kern-5.truept\to\kern-1.truept}}

\def\={{ \; \equiv \; }}		
\ifnum\mgnf=0
    \def\openone{\leavevmode\hbox{\ninerm 1\kern-3.3pt\tenrm1}}%
\fi
\ifnum\mgnf=1
     \def\openone{\leavevmode\hbox{\ninerm 1\kern-3.6pt\tenrm1}}%
\fi

\def\2{{1\over2}}

\def\igb{
    \mathop{\raise4.pt\hbox{\vrule height0.2pt depth0.2pt width6.pt}
    \kern0.3pt\kern-9pt\int}}

\def\st{\scriptscriptstyle}
\let\\=\noindent
\def\*{\vskip0.5truecm}
\def\?{\vskip0.75truecm}
\def\item#1{\vskip0.1truecm\parindent=0pt\par\setbox0=\hbox{#1}
     \hangindent\ifdim\wd0>0.6cm 0.6cm\else\wd0\fi\hangafter 1 #1 \parindent=\stdindent}

\def\annota#1#2{{\footnote{${}^#1$}{\ottopunti
\ifnum\mgnf=0\baselineskip=9pt\normalbaselineskip=9pt
\else\baselineskip=18pt\normalbaselineskip=18pt\fi
\parindent=0pt#2\vfill}}}
\def\annotano#1{\annota{\number\noteno}{#1}\advance\noteno by 1}


\def\ie{\hbox{\sl i.e.\ }}

\def\qed{\hfill\break\nobreak\vbox{\vglue.25truecm\line{\hfill\raise1pt 
          \hbox{\vrule height9pt width5pt depth0pt}}}\vglue.25truecm}


\def\gint(#1)(#2)(#3){{\cal D}#1^{#2}\,e^{(#1^{#2+},#3#1^{#2-})}}

  \def\V0{{\bf 0}}    \def\tt{{\bf t}}

\def\BB{{\cal B}}

\def\SS{{\cal S}}


\ifnum\cind=1
\def\prtindex#1{\immediate\write\indiceout{\string\parte{#1}{\the\pageno}}}
\def\capindex#1#2{\immediate\write\indiceout{\string\capitolo{#1}{#2}{\the\pageno}}}
\def\parindex#1#2{\immediate\write\indiceout{\string\paragrafo{#1}{#2}{\the\pageno}}}
\def\subindex#1#2{\immediate\write\indiceout{\string\sparagrafo{#1}{#2}{\the\pageno}}}
\def\appindex#1#2{\immediate\write\indiceout{\string\appendice{#1}{#2}{\the\pageno}}}
\def\paraindex#1#2{\immediate\write\indiceout{\string\paragrafoapp{#1}{#2}{\the\pageno}}}
\def\subaindex#1#2{\immediate\write\indiceout{\string\sparagrafoapp{#1}{#2}{\the\pageno}}}
\def\bibindex#1{\immediate\write\indiceout{\string\bibliografia{#1}{Bibliografia}{\the\pageno}}}
\def\preindex#1{\immediate\write\indiceout{\string\premessa{#1}{\the\pageno}}}
\else
\def\prtindex#1{}
\def\capindex#1#2{}
\def\parindex#1#2{}
\def\subindex#1#2{}
\def\appindex#1#2{}
\def\paraindex#1#2{}
\def\subaindex#1#2{}
\def\bibindex#1{}
\def\preindex#1{}
\fi

\def\leaderfill{\leaders\hbox to 1em{\hss . \hss} \hfill }


\newdimen\capsalto \capsalto=0pt
\newdimen\parsalto \parsalto=20pt
\newdimen\sparsalto \sparsalto=30pt
\newdimen\tratitoloepagina \tratitoloepagina=2\parsalto
\def\aboveparteskip{\bigskip \bigskip}
\def\belowparteskip{\medskip \medskip}
\def\abovecapitskip{\bigskip}
\def\belowcapitskip{\medskip}
\def\belowparskip{\smallskip}
%


\def\parte#1#2{
   \9{\immediate\write16
      {#1     pag.\numeropag{#2} }}
   \aboveparteskip 
   \noindent 
   {\ftitolo #1} 
   \hfill {\ftitolo \numeropag{#2}}\par
   \belowparteskip
   }


\def\premessa#1#2{
   \9{\immediate\write16
      {#1     pag.\numeropag{#2} }}
   \abovecapitskip 
   \noindent 
   {\it #1} 
   \hfill {\rm \numeropag{#2}}\par
   \belowcapitskip
   }


\def\bibliografia#1#2#3{
  \ifnum\stile=1
   \9{\immediate\write16
      {Bibliografia    pag.\numeropag{#3} }}
   \belowcapitskip
   \noindent 
   {\bf Bibliografia} 
   \hfill {\bf \numeropag{#3}}\par
  \else
    \paragrafo{#1}{References}{#3}
\fi
   }


\newdimen\newstrutboxheight
\newstrutboxheight=\baselineskip
\advance\newstrutboxheight by -\dp\strutbox
\newdimen\newstrutboxdepth
\newstrutboxdepth=\dp\strutbox
\newbox\newstrutbox
\setbox\newstrutbox = \hbox{\vrule 
   height \newstrutboxheight 
   width 0pt 
   depth \newstrutboxdepth 
   }
\def\newstrut
   {\relax \ifmmode \copy \newstrutbox \else \unhcopy \newstrutbox \fi}
%
%
\vfuzz=3.5pt
%
%
\newdimen\indexsize \indexsize=\hsize
\advance \indexsize by -\tratitoloepagina
\newdimen\dummy
\newbox\parnum
\newbox\parbody
\newbox\parpage
%

%

\def\mastercap#1#2#3#4#5{
   \9{\immediate\write16
      {Cap. #3:#4     pag.\numeropag{#5} }}
   \abovecapitskip
   \setbox\parnum=\hbox {\kern#1\newstrut{#2}
			{\bf Capitolo~\number#3.}~}
   \dummy=\indexsize
   \advance\indexsize by -\wd\parnum
   \setbox\parbody=\vbox {
      \hsize = \indexsize \noindent \newstrut 
      {\bf #4}
      \newstrut \hss}
   \indexsize=\dummy
   \setbox\parnum=\vbox to \ht\parbody {
      \box\parnum
      \vfill 
      }
   \setbox\parpage = \hbox to \tratitoloepagina {
      \hss {\bf \numeropag{#5}}}
   \noindent \box\parnum\box\parbody\box\parpage\par
   \belowcapitskip
   }
\def\capitolo#1#2#3{\mastercap{\capsalto}{}{#1}{#2}{#3}}
%


%
\def\masterapp#1#2#3#4#5{
   \9{\immediate\write16
      {App. #3:#4     pag.\numeropag{#5} }}
   \abovecapitskip
   \setbox\parnum=\hbox {\kern#1\newstrut{#2}
			{\bf Appendice~A\number#3:}~}
   \dummy=\indexsize
   \advance\indexsize by -\wd\parnum
   \setbox\parbody=\vbox {
      \hsize = \indexsize \noindent \newstrut 
      {\bf #4}
      \newstrut \hss}
   \indexsize=\dummy
   \setbox\parnum=\vbox to \ht\parbody {
      \box\parnum
      \vfill 
      }
   \setbox\parpage = \hbox to \tratitoloepagina {
      \hss {\bf \numeropag{#5}}}
   \noindent \box\parnum\box\parbody\box\parpage\par
   \belowcapitskip
   }
\def\appendice#1#2#3{\masterapp{\capsalto}{}{#1}{#2}{#3}}
%

%

\def\masterpar#1#2#3#4#5{
   \9{\immediate\write16
      {par. #3:#4     pag.\numeropag{#5} }}
   \setbox\parnum=\hbox {\kern#1\newstrut{#2}\number#3.~}
   \dummy=\indexsize
   \advance\indexsize by -\wd\parnum
   \setbox\parbody=\vbox {
      \hsize = \indexsize \noindent \newstrut 
      #4
      \newstrut \hss}
   \indexsize=\dummy
   \setbox\parnum=\vbox to \ht\parbody {
      \box\parnum
      \vfill 
      }
   \setbox\parpage = \hbox to \tratitoloepagina {
      \hss \numeropag{#5}}
   \noindent \box\parnum\box\parbody\box\parpage\par
   \belowparskip
   }
\def\paragrafo#1#2#3{\masterpar{\parsalto}{}{#1}{#2}{#3}}
\def\sparagrafo#1#2#3{\masterpar{\sparsalto}{}{#1}{#2}{#3}}
%


%
\def\masterpara#1#2#3#4#5{
   \9{\immediate\write16
      {par. #3:#4     pag.\numeropag{#5} }}
   \setbox\parnum=\hbox {\kern#1\newstrut{#2}A\number#3.~}
   \dummy=\indexsize
   \advance\indexsize by -\wd\parnum
   \setbox\parbody=\vbox {
      \hsize = \indexsize \noindent \newstrut 
      #4
      \newstrut \hss}
   \indexsize=\dummy
   \setbox\parnum=\vbox to \ht\parbody {
      \box\parnum
      \vfill 
      }
   \setbox\parpage = \hbox to \tratitoloepagina {
      \hss \numeropag{#5}}
   \noindent \box\parnum\box\parbody\box\parpage\par
   \belowparskip
   }
\def\paragrafoapp#1#2#3{\masterpara{\parsalto}{}{#1}{#2}{#3}}
\def\sparagrafoapp#1#2#3{\masterpara{\sparsalto}{}{#1}{#2}{#3}}
%


\ifnum\stile=1

\def\newcap#1{\setcap{#1}
\vskip2.truecm\advance\numsec by 1
\\{\ftitolo \numero\numsec. #1}
\capindex{\numero\numsec}{#1}
\vskip1.truecm\numfor=1\pgn=1\numpar=1\numteo=1\numlem=1
}

\def\newapp#1{\setcap{#1}
\vskip2.truecm\advance\numsec by 1
\\{\ftitolo A\numero\numsec. #1}
\appindex{A\numero\numsec}{#1}
\vskip1.truecm\numfor=1\pgn=1\numpar=1\numteo=1\numlem=1
}

\def\newpar#1{
\vskip1.truecm
\vbox{
\\{\bf \numero\numsec.\numero\numpar. #1}
\parindex{\numero\numsec.\numero\numpar}{#1}
\*{}}
\nobreak
\advance\numpar by 1
}

\def\newpara#1{
\vskip1.truecm
\vbox{
\\{\bf A\numero\numsec.\numero\numpar. #1}
\paraindex{\numero\numsec.\numero\numpar}{#1}
\*{}}
\nobreak
\advance\numpar by 1
}

\else

\def\newsec#1{\vskip1.truecm
\advance\numsec by 1
\vbox{
\\{\bf \numero\numsec. #1}
\parindex{\numero\numsec}{#1}
\*{}}\numfor=1\pgn=1\numpar=1\numteo=1\numlem=1
\nobreak
}

\def\newsubsect#1{
\vskip1.truecm
\vbox{
\\{\it \numero\numsec.\numero\numpar. #1}
\parindex{\numero\numsec.\numero\numpar}{#1}
\*{}}
\nobreak
\advance\numpar by 1
}

\def\newapp#1{\vskip1.truecm
\advance\numsec by 1
\vbox{
\\{\bf A\numero\numsec. #1}
\appindex{A\numero\numsec}{#1}
\*{}}\numfor=1\pgn=1\numpar=1\numteo=1\numlem=1
}

\def\biblio{\vskip1.truecm
\vbox{
\\{\bf References.}\*{}
\bibindex{{}}}\nobreak\makebiblio
}

\fi


\newread\indicein
\newwrite\indiceout

\def\faindice{
\openin\indicein=\jobname.ind
\ifeof\indicein\relax\else{
\ifnum\stile=1
  \pagenumbersind
  \pageno=-1
  \setind
  \null
  \vskip 2.truecm
  \\{\ftitolo Indice}
  \vskip 1.truecm
  \parskip = 0pt
  \input \jobname.ind
  \ppaginan
\else
\\{\bf Index}
\*{}
 \input \jobname.ind
\fi}\fi
\closein\indicein
\def\nomeindice{\jobname.ind}
\immediate\openout \indiceout = \nomeindice
}


\newwrite\bib
\immediate\openout\bib=\jobname.bib
\global\newcount\bibex
\bibex=0
\def\verabib{\number\bibex}

\ifnum\tipobib=0
\def\cita#1{\expandafter\ifx\csname c#1\endcsname\relax
\hbox{$\clubsuit$}#1\write16{Manca #1 !}%
\else\csname c#1\endcsname\fi}
\def\rife#1#2#3{\immediate\write\bib{\string\raf{#2}{#3}{#1}}
\immediate\write15{\string\C(#1){[#2]}}
\setbox199=\hbox{#2}\ifnum\maxit < \wd199 \maxit=\wd199\fi}
\else
\def\cita#1{%
\expandafter\ifx\csname d#1\endcsname\relax%
\expandafter\ifx\csname c#1\endcsname\relax%
\hbox{$\clubsuit$}#1\write16{Manca #1 !}%
\else\probib(ref. numero )(#1)%
\csname c#1\endcsname%
\fi\else\csname d#1\endcsname\fi}%
\def\rife#1#2#3{\immediate\write15{\string\Cp(#1){%
\string\immediate\string\write\string\bib{\string\string\string\raf%
{\string\verabib}{#3}{#1}}%
\string\Cn(#1){[\string\verabib]}%
\string\CCc(#1)%
}}}%
\fi

\def\Cn(#1)#2{\expandafter\xdef\csname d#1\endcsname{#2}}
\def\CCc(#1){\csname d#1\endcsname}
\def\probib(#1)(#2){\global\advance\bibex+1%
\9{\immediate\write16{#1\verabib => #2}}%
}

\def\C(#1)#2{\SIA c,#1,{#2}}
\def\Cp(#1)#2{\SIAnx c,#1,{#2}}

\def\SIAnx #1,#2,#3 {\senondefinito{#1#2}%
\expandafter\def\csname#1#2\endcsname{#3}\else%
\write16{???? ma #1,#2 e' gia' stato definito !!!!}\fi}

\bibskip=10truept
\def\hboxto{\hbox to}

\catcode`\{=12\catcode`\}=12
\catcode`\<=1\catcode`\>=2
\immediate\write\bib<
	\string\halign{\string\hboxto \string\maxit%
	{\string #\string\hfill}&%
        \string\vtop{\string\parindent=0pt\string\advance\string\hsize%
	by -1.55truecm%
	\string#\string\vskip \bibskip
	}\string\cr%
>
\catcode`\{=1\catcode`\}=2
\catcode`\<=12\catcode`\>=12

\def\raf#1#2#3{\ifnum \bz=0 [#1]&#2 \cr\else
\llap{${}_{\rm #3}$}[#1]&#2\cr\fi}

\newread\bibin

\catcode`\{=12\catcode`\}=12
\catcode`\<=1\catcode`\>=2
\def\chiudibib<
\catcode`\{=12\catcode`\}=12
\catcode`\<=1\catcode`\>=2
\immediate\write\bib<}>
\catcode`\{=1\catcode`\}=2
\catcode`\<=12\catcode`\>=12
>
\catcode`\{=1\catcode`\}=2
\catcode`\<=12\catcode`\>=12

\def\makebiblio{
\ifnum\tipobib=0
\advance \maxit by 10pt
\else
\maxit=1.truecm
\fi
\chiudibib
\immediate \closeout\bib
\openin\bibin=\jobname.bib
\ifeof\bibin\relax\else
\raggedbottom
\input \jobname.bib
\fi
}

\openin13=#1.aux \ifeof13 \relax \else
\input #1.aux \closein13\fi
\openin14=\jobname.aux \ifeof14 \relax \else
\input \jobname.aux \closein14 \fi
\immediate\openout15=\jobname.aux

\def\V#1{\underline{#1}}

\def\normalbaselines{\baselineskip=20pt\lineskip=3pt\lineskiplimit=3pt}

\def\normalbaselines{\baselineskip=20pt\lineskip=3pt\lineskiplimit=3pt}

\def\bE{{\bf E}}
\def\bv{{\bf v}}
\def\bq{{\bf q}}
\def\bV{{\bf V}}
\def\bQ{{\bf Q}}
\def\bR{{\bf R}}
\def\bTheta{{\Theta}}
\def\bJ{{\bf J}}

\def\bj{{\bf j}}

\def\bk{{\bf k}}
\def\bn{{\bf n}}

\large\hsize=6.5truein\hoffset=-0.0in

\titolo{Properties of Stationary Nonequilibrium States in the Thermostatted Periodic Lorentz Gas II: The many point particles system}

\autore{F. Bonetto}{School of Mathematics, Institute for Advanced Study, Princeton, NJ 08540. Email: {\tt bonetto@ias.edu}}
\autore{D. Daems}{Center for Nonlinear Phenomena and Complex Systems, Universit\'e Libre de Bruxelles, 1050 Brussels, Belgium. Email: {\tt ddaems@ulb.ac.be}}
\autore{J.L. Lebowitz}{School of Mathematics, Institute for Advanced Study, Princeton, NJ 08540.
\hfill\break Email: {\tt Lebowitz@math.rutgers.edu}}
\autore{V. Ricci}{Dipartimento di Matematica, 
Universit\`a ``La Sapienza'', Piazzale Aldo Moro n.5, 00185 Roma,
Italy. \hfill\break Email: {\tt Valeria.Ricci@roma1.infn.it} or {\tt ricci@mat.uniroma1.it}}

\abstract{We study the stationary nonequilibrium states 
of $N$ point particles moving under the influence of an electric field
${\bf E}$ among fixed obstacles (discs) in a two dimensional torus. The total
kinetic energy of the system
is kept constant through a Gaussian thermostat which
produces a velocity dependent mean field interaction between the
particles.  The current and the particle distribution functions are
obtained numerically and compared for small $|\bE|$ with analytic
solutions of a Boltzmann type equation obtained by treating the
collisions with the obstacles as random independent 
scatterings.  The agreement is
surprisingly good for both small and large $N$. The latter system in turn
agrees with a self consistent one particle evolution expected to 
hold in the $N\to\infty$ limit.}

\parole{Lorentz Gas, Gaussian Thermostat, Electrical Current, Steady state}

\prima

\newsec{Introduction}

In this note we continue our study of the stationary nonequilibrium
states (SNS) of current carrying thermostatted systems.  In part I 
\cita{BDL} we
described extensive numerical and analytical investigations of the
dependence of the current on the electric field for a model single
particle system introduced in \cita{MH} and previously studied in
\cita{CELS}.  Here we study a generalization of that model to
$N$ particles introduced in \cita{BGG}.  The particles, which have
unit mass, move among a fixed periodic array of discs in a two
dimensional square $\Lambda$ with periodic boundary 
conditions, see Fig. 1. They are acted on by an external
(electric) field $\bE$ parallel to the $x$-axis and by a ``Gaussian
thermostat''.  (The discs are located so that there is a finite
horizon, \ie there is a maximum distance a particle can move before
hitting a disc or obstacle).

\eqfig{600pt}{200pt}{
\ins{298pt}{24pt}{${\nota 2L}$}
\ins{296pt}{57pt}{${\nota \bv_1}$}
\ins{252pt}{94pt}{${\nota \bv_2}$}
\ins{335pt}{107pt}{${\nota \bv_3}$}
\ins{292pt}{193pt}{${\nota \bE}$}
\ins{250pt}{76pt}{${\nota R_2}$}
\ins{295pt}{120pt}{${\nota R_1}$}
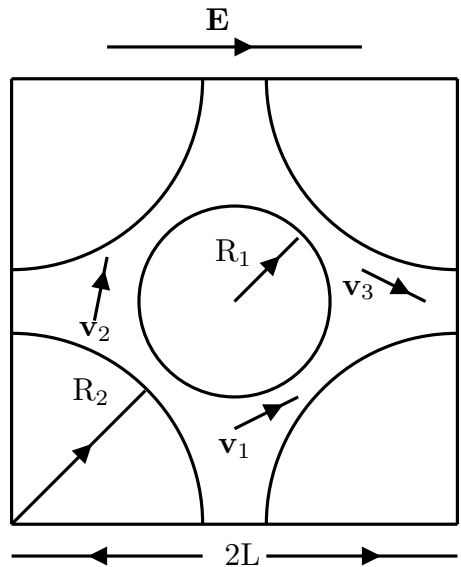
}{x1}{General billiard structure with discs of radius $R_1$ and $R_2$
in a periodic box with side length $2L$, $N=3$ particles are shown.}  {fig1}
The equations of motion describing the time evolution of the positions
$\bq_i$ and velocities ${\bf v}_i, i=1,...,N$, are:

$$\left\{\eqalign{\dot\bq_i=&\bv_i\qquad
\bq_i=(q_{i,x},q_{i,y})\in\Lambda'\cr
\dot\bv_i=&\bE-\alpha(\bJ,U)\bv_i+F_{obs}(\bq_i)\cr}\right.\Eq(dyn1)$$
where

$$\alpha(\bJ,U)={\bJ\cdot\bE \over U},
\qquad\bJ={1\over N}\sum_{i=1}^N\bv_i, \qquad U={1\over
N}\sum_{i=1}^N\bv_i^2\Eq(dyn3)$$
Here $\Lambda'=\Lambda\backslash{\cal D}$, with ${\cal D}$ the region
occupied by the discs (obstacles) and $F_{obs}$ represents the elastic
scattering which takes place at the surface of the obstacles. The
purpose of the Gaussian thermostat, represented by the term
$\alpha(\bJ,U){\bf v}$ in eq.\equ(dyn1), is to maintain the total
kinetic energy $1/2\sum_{i=1}^N\bv_i^2$ constant, \ie $U=v_0^2$. It
also has the effect of making the flow $\Phi_t$ generated by
eq.\equ(dyn1) on the $(4N-1)$ dimensional energy surface non
Hamiltonian when ${\bf E}\not=0$. In fact the phase space volume
contraction rate is given by
$\sigma(X)=-(2N-1)\alpha(\bJ,U)$. Another effect of the thermostat
is to effectively couple all the particles in a mean field way,
$\alpha({\bf J},U)$, depending only on the total momentum of the
particles.  Note that this is the only coupling between the particles
in this system.

The change of variables, $\bq_i\to\bq_i/L$, $\bv_i\to\bv_i/v_0$, $t\to
tv_0/L$ and $\bE\to\bE L/v_0^2$, where $2L$ is the length of the box,
leaves eq.\equ(dyn1) unchanged, so that the motion of the system takes
place on $\SS_N=(\Lambda')^N\times S_N$, where
$S_N=\{\bv_i|\sum_{i=1}^N\bv_i^2=N\}$. We shall denote by $X\in\SS_N$ a
point in the phase space of the system.  In these units we took $U=1$,
$R_1=0.39$, $R_2=0.79$, and $\Lambda$ is thetorus of 
side 2\annotano{See \cita{BDL} for an explanation of these values.}. 

Our main interest is in the SNS of this model system. To
be more precise let $\mu_0(dX,N)=\rho_0(X;N)dX$ be an initial measure
symmetric in the $\{\bq_i,\bv_i\}$ and absolutely continuous with
respect to the Liouville volume $dX$ projected on $\SS_N$.  The time
evolved measure $\mu_t(dX,\bE;N)$ is still absolutely continuous
with respect to the Liouville measure with  
density $\rho_t(X,\bE;N)$ for any fixed time $t$.
The SNS is expected to be
described by an SRB measure $\mu^+(dX,\bE;N)$, given by the weak limit,  
as $t\longrightarrow\infty$, of
$\mu_t(dX,\bE;N)$, when it exists. This limit measure is in general not 
absolutely
continuous with respect to the
Liouville measure, due to the phase space volume contraction \cita{Ru}, \cita{Ru3}. 
The existence of
such a limit was proven, for $N=1$ and
 $|\bE|\in[0,E_0]$
($E_0$ small)
in \cita{CELS}, but no such result is available for
$N\ge 2$, because of the lack of uniform
hyperbolicity for the zero field system.  On the other hand our
computer simulations of the dynamics, for $N$ ranging from 1 to 50 and
$E$ from $0.04$ to $1.0$, strongly support the belief that there exists
a unique limiting measure $\mu^+(dX,\bE;N)$ up to quite large values
of $|\bE|$, say $|\bE|=E \le 1$.  
We expect however that the projection of
$\mu^+(dX,\bE;N)$ on the one particle phase space
$\Lambda'\times\Omega_{(N)}$, where $\Omega_{(N)}$ is the ball
$|\bv|\le\sqrt{N}$, will yield a one particle density
$f^+(\bq,\bv,\bE;N)$ absolutely continuous with respect to 
$d\bq d\bv$; this is proven, for instance, for coupled Arnold's cat maps
\cita{BKL}). 

To obtain information about $f^+$ we considered first the case of weak
fields.  It is tempting to think that for $E\longrightarrow 0$ the
singular set on which $\mu^+$ is concentrated will be spread out more
or less uniformly on $\SS_N$ so that $\mu^+$ will approach weakly the
microcanonical measure on the energy surface $\SS_N$: this measure is
certainly invariant for the dynamics at $E=0$. If this were the case
then $f^+(\bq,\bv,\bE;N)$ would approach, as $E\to 0$, the equilibrium
one particle density obtained from the projection of the
microcanonical measure: for large $N$ this would be close to the
Maxwellian distribution with unit variance \annotano{Note that for 
large $N$ the Maxwell distribution is
typical for points on the energy surface, \ie the set $\BB$ on $\SS_N$
for which $f^+$ is not a Maxwellian has measure 0 (w.r.t. $dX$). Of
course since $\mu^{+}$ is singular w.r.t. $dX$ this need not to be the
case here.}. We ran computer
simulations for values of the field between $0.04$ and $0.12$ and
$N=2,5$ and 50. In all cases we found a one particle distribution that
is far from the projection of the microcanonical distribution.
Furthermore this distribution appeared to have only very slight
dependence on $E$ for those values of the field; so it appears that
there is a well defined limit of $f^+(\bq,\bv,\bE; N)$ as $E\to 0$,
and that this limit is {\it not} the projection of the microcanonical
measure: there are correlations between the velocities of the
particles induced by the field, beyond those corresponding to the
energy constraint, which remain when $E\longrightarrow 0$.

This deviation from the microcanonical distribution is reflected also
in the behavior of the average current per particle in the steady
state, given by $\bj(\bE,N)=\int\bv f^+(\bq,\bv,\bE;N)d\bq d\bv$ as
$E\to 0$. We studied $\bj(\bE,N)$ numerically as a function of $\bE$
and $N$, see Fig.\equ(fig2) and Fig.\equ(fig3). In the following we
will always assume that the electric field is along the positive
$x$-axis, $\bE=E{\bf 1}_x$. This implies that the $y$ component of
$\bj(\bE,N)$ is zero for symmetry reason. We will denote the $x$
component of the current by $j(E,N)$ and call $\k(E,N)=j(E,N)/E$ the
conductivity.  The dependence on $N$ for $E\to 0$ should be given by
the Green-Kubo formula for the zero field conductivity when the
dynamics of the particles are independent. A straightforward
computation then shows that the zero field conductivity of the $N$
particles is:

$$\k(0,N)=C_N(0)\k(0,1)\Eq(GK)$$
with $\k(0,1)$ given by the diffusion constant of Bunimovich and Sinai
\cita{BS} and

$$C_N(0)=\int {1\over |\bv|}f^+(\bq,\bv,0;N)d\bq d\bv\Eq(gk)$$
For the microcanonical distribution we easily find:

$$C_N(0)=\sqrt{\pi}\left(1-{3\over 8N}+O\left(N^{-2}\right)\right)\Eq(mic)$$
which is inconsistent with our data although the form of the
dependence on $N$ appear to be similar, see sect. 2.1.

Let us consider now the behavior of our model system in the limit
$N\to\infty$. As the particles interact only through their average
velocity $\bJ (X(t))$ it seems reasonable to expect that, for
$N\to\infty$, ${\bJ}$ will stop fluctuating,
\ie that 
for ``well behaved'' initial distributions \cita{BH}, \cita{Lan}, \cita{S}
$$
\bJ (X(t))\longrightarrow
{\bj}_t =\int \bv f_t(\bv,\bE)d\bv
\Eq(cur)
$$
where $f_t(\bv,\bE)=\lim_{N\longrightarrow\infty}
f_t(\bv,\bE;N)$.  If this were true in a sufficiently strong
sense it would lead to an autonomous Vlasov type equation
\cita{BH}, \cita{Lan}, \cita{S} for $f_t$ where $\dot{\bv}$ would be
computed self consistently from the (irreversible) dynamics
\annotano{The dynamics \equ(dyn1) is reversible in the sense that if
$T_t X$ is a solution then $T_t R T_t X=RX$, where $R$ reverses all
velocities.}

$${\dot \bv} = \bE - \lambda(t)\bv + {\bf F}_{obs}(\bq)\Eq(vi)$$
with $\lambda(t)=\bE\cdot{\bj}_{t}$.  The difficulty with proving this
behavior, as compared to the \cita{BH} case, is that trajectory
$X(t)$ and thus also $\bJ (X(t))$ is not smooth for finite $t$.  The
problems are compounded when we consider the $t\longrightarrow\infty$
limit corresponding to the SNS.

Based on numerical evidence we nevertheless believe that
$$\lim_{N\to\infty} f^+(\bv,\bE;N)=\hat f^+(\bv,\bE)\equiv \lim_{t\to\infty}\hat
f_t(\bv,\bE)\Eq(equv)$$
where $\hat f_t(\bv,\bE)$ is the solution of the Vlasov equation with
a force given by the right hand side of \equ(vi), and we define for a
given function $g$

$$g(\bv)=\int_{\Lambda'}g(\bq,\bv)d\bq \ .$$

The integration over $\bq$ is necessary, or at least desirable, since
we expect the $t\to\infty$ limit of $\hat f_t(\bq,\bv,\bE)$ to be
singular with respect to $d\bq d\bv$ as is the $N=1$ reversible system
\equ(dyn1). Its projection on the velocity is however expected to be
absolutely continuous with respect to $d\bv$ \cita{CELS}\cita{BKL}
. Eq\equ(equv) is thus a form of the law of large numbers which should
hold for smooth $\rho_0 (X,\bE;N)$.  Something like this was in fact
proven by Ruelle for the stationary state under some hypotheses on the
thermostatted dynamics \cita{Ru2}.
To make contact with Ruelle's theorem it is convenient to think of
$\Lambda_N$ as a torus of length $2LN$ along the $y$-axis
(perpendicular to $\bE$) and length $2L$ along the $x$-axis. This does
not change the dynamics.

To get some analytical handle on the form of the reduced distributions
in the SNS we investigated a model system in which the deterministic
collisions with the obstacles are replaced by a stochastic process in
which particle velocities get their orientations changed at random
times, independent for each particle.  This yields a Markov process
which replaces the continuity equation for $\rho_t(X,E;N)$ by a linear
Boltzmann-like equation, see \cita{VB1}.  We can write either of these 
equations in the symbolic form:

$${\partial\over \partial t}\rho_t(\bQ,\bV)+
\sum_i{\partial\over \partial \bq_i}\left\{\bv_i \rho_t(\bQ,\bV)\right\}+
\sum_i{\partial\over \partial \bv_i}\left\{\left[\bE-
\alpha(\bV)\bv_i\right] \rho_t(\bQ,\bV)\right\}=
\left({\partial \rho_t\over \partial t}\right)_{{\rm
coll}}, \Eq(bltz)$$
where we have set $X = ({\bf Q},{\bf V})$ (and dropped the explicit
dependence on $\bf E$ and $N$). The term on the right hand
$\left({\partial \rho_t\over\partial t}\right)_{{\rm coll}}$
represents either the effect of deterministic collisions with the
obstacles as given by \equ(dyn1) or a collision operator independent
of $\bQ$, see \equ(coll).  A similar ansatz for the irreversible
dynamics \equ(vi) leads to a Boltzmann-Vlasov equation for the one
particle distribution.  These equations can be solved analytically as
a power series in $E$ and/or numerically.  This is described in
sect. 3.

In sect. 4 we compare some of the moments, including the current, of
the deterministic distribution $f^{+}(\bv,\bE;N)$ with those of the
stochastic one.  We find surprisingly good agreement once the mean
free path appearing in the Boltzmann-like equations is properly
interpreted, see sec. 4.2. We note however that a direct computation
of the distribution of free paths in the dynamical system \equ(dyn1)
shows that it is far from being exponential, which is the basic
assumption of the Markov process. We therefore have no real
explanation for the observed good agreement.  We only note that some
features of the stationary state appear rather robust with respect to
the collision processes with the ``obstacles'', yielding similar
results for different distributions for the free path. In sect. 5 we
discuss some general questions about the relation between this
thermostatted model and the Drude model of electrical conduction in
metals \cita{DR}.

\newsec{Numerical results} 

Eq. \equ(dyn1) can be solved in terms of quadratures between
collisions with the obstacles so the simulation consists mainly in
computing the times of successive collisions. At each collision there
is an instantaneous change in the velocity of the colliding particle
and consequently also in the current $\bJ$ and thus in the
thermostatted force acting on each particle. Assuming that the system
is ergodic we can obtain information about the SNS from time averages
over a single trajectory. In practice we used a few initial states and
found a behavior consistent with this assumption. The relative
simplicity of the dynamics enabled us to get fairly accurate results
even for 50 particles with relatively small computing power. Our
simulation were carried out on a Pentium PC. Error bars are computed
by doubling the range of the fluctuations of the time average over the
interval $[0.9T,T]$ where $T$ is the total number of collisions
computed. After the change of variables described after \equ(dyn3) all 
quantities appearing in the graphs are adimensional. 

\newsubsect{The current}

Let $\bj(\bE,N)$ be the average current in the steady state $\mu^{+}$,

$$\bj(\bE,N)=\langle \bJ \rangle_{\mu^+}= 
\int\bv f^+(\bv,\bE, N)d\bv\ .\Eq(j)$$
with $\bJ$ defined in \equ(dyn3).  As already noted, in all our
computations the electric field is along the positive $x$-axis, ${\bf
E}=E {\bf 1_x}$, all densities are normalized and $j(E,N)$ is the
$x$-component of the current defined in eq.\equ(j).

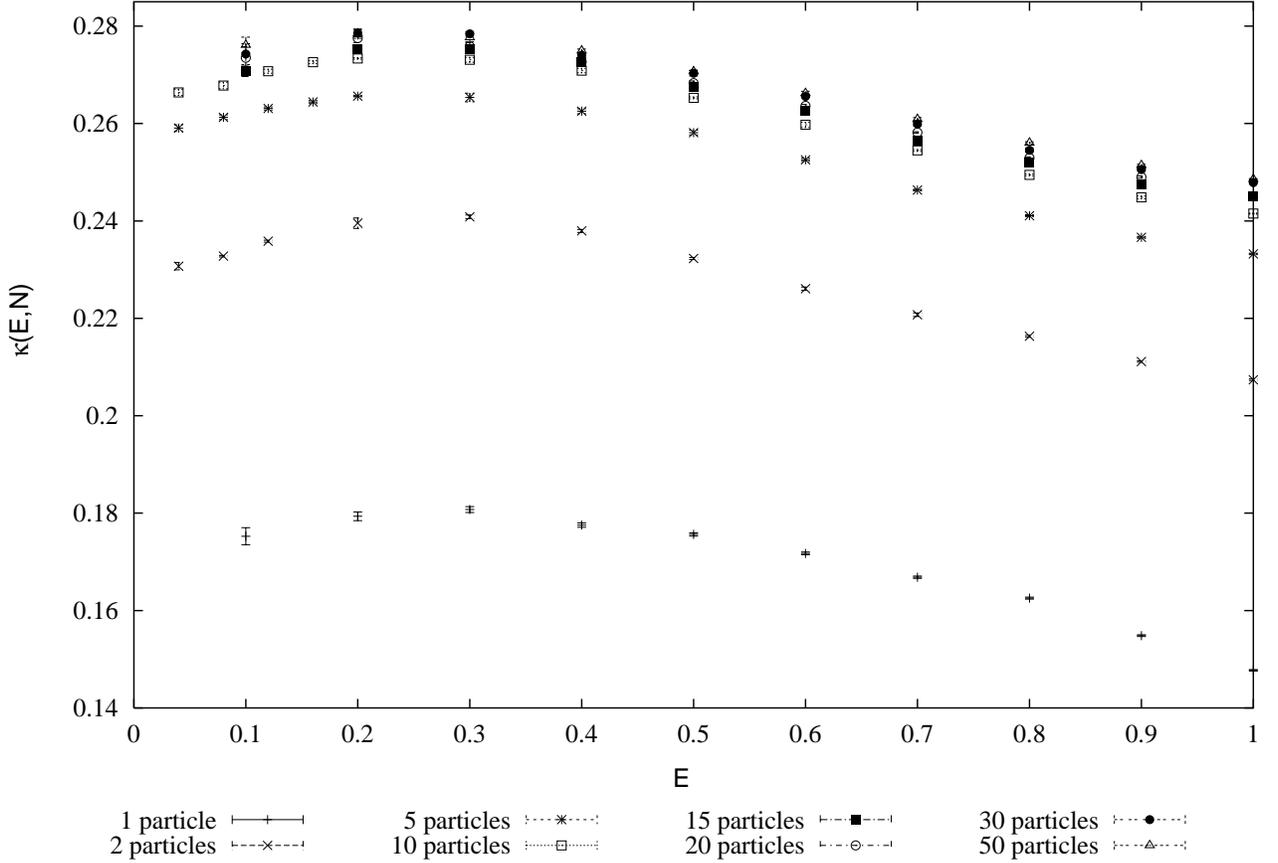
\gnuins fig2 fig2 {Conductivity $\k(E,N)$ as a function of $E$ for 
different $N$.}

In Fig.\equ(fig2) we plot the conductivity $\kappa(E,N)=j(E,N)/E$ as a
function of the field for different numbers of particles,
$N$=1,2,10,15,20,30 and 50. The averages were computed by running
simulations in which the total number of collisions with the obstacles
varied from $10^9$ for $N=1$ to $10^8$ for $N=50$.

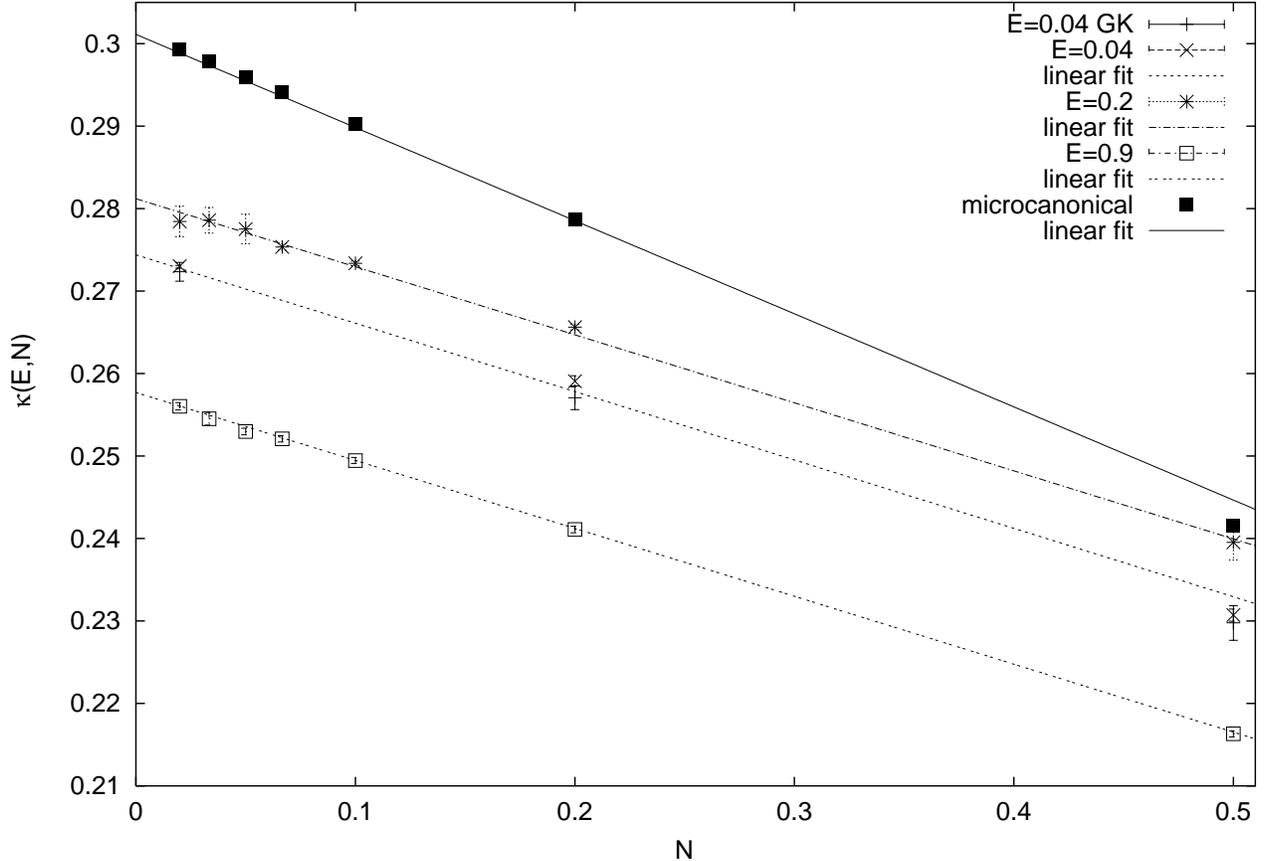
\gnuins fig3 fig3 {$\k(E,N)$ as a function of $N^{-1}$ for 
different $E$. Also plotted is the conductivity obtained from
eq.\equ(GK) using the actual distribution function, see next section, for 
$E=0.04$ and compared with the value obtained by a direct simulation 
at the same field. Finally the highest line represents the conductivity 
obtained from eq.\equ(GK) using a microcanonical hypothesis.}

We note that for very small fields the interaction
among the particles is very small so that the invariant distribution
is reached only after a very long transient time.

Furthermore, although the current goes to 0 as $E\to 0$, the
fluctuations in the current are almost independent of $E$ so that
longer and longer simulations are required in order to distinguish the
average from the fluctuations when $E\to 0$. For
$N=2,5$ and $10$ we checked whether ${d \k(E,N)\over dE}\to 0$ as $E\to
0$, as required by the symmetry of the problem if $\k(E,N)$ is
differentiable at 0. While the results are not definitive they are
consistent with such behavior.

In Fig. \equ(fig3) we plot the conductivity as a function of $1/N$ for
a few selected values of the field. As can be seen there the behavior
of $\kappa(E,N)$ can be well fitted for $N>2$ by the following formula
which is the analogous of eq. \equ(GK) with $C_N(0)$ given by
\equ(mic) for $E \neq 0$:
$\kappa(E,N)=\tilde\kappa(E)+c/N$ with
$\tilde\kappa(E)=\lim_{N\to\infty} \kappa(E,N)$ and $c$ independent
from $E$, at least within the accuracy of our computation. (The value
of $\k(E,1)$ is about 15-20\% lower than that given by the
formula, depending on $E$).  For $E=0.04$ we have the value of the
conductivity for $N=2,5$ and 50 as well as the distribution
$f^+(\bv,E;N)$. We can therefore check directly
eq.\equ(gk) for $E \neq 0$. Fig. \equ(fig3) contains both the values obtained
directly and those obtained from eq.\equ(gk) for $E=0.04$. The agreement is clearly
very good. Finally plotted in Fig.\equ(fig3) is the value of the
conductivity at zero field obtained from eq.\equ(mic), \ie assuming
that the invariant distribution is microcanonical. Although this
assumption is inconsistent with the actual numerical
data, the behavior is qualitatively similar.

The smoothness, or rather the lack of smoothness, of the current as a
function of $E$ for $N=1$ was extensively discussed in \cita{BDL} and
related there to the discontinuities of the collision map. The data we
have for $N \ge 2$ are insufficient to address this question.  However
it is expected that the stationary current will be smoother than it is
in the one particle case, since it is averaged over all particles.

\newsubsect{Distribution functions}

To study the space independent part of the one particle density
function, $f^+(\bv,E;N)$, it is convenient to switch to the variables
$r=\vert \bv\vert
\in [0,\sqrt{N}]$ and $\theta \in [-\pi,\pi]$ the angle between the
velocity ${\bf v}$ and the $x$-axis.  Expanding $f^{+}(\bv,E,N)$ in a
Fourier series in $\theta$, we have
 
$$f^{+}(\bv,E;N)= \sum_{k=0}^{\infty}\psi_k(r,E;N) \cos k\theta \
, \Eq(esp)$$  
where only terms in $\cos k\theta$ appear due to the symmetry of the
problem. Note that $2\pi r\psi_0(r,E;N)$ 
is the stationary probability density
for
the modulus of $\bv$ while

$$j(E,N)=\pi\int_0^{\sqrt{N}}dr\, r^2 \psi_1(r,E;N) \ .\Eq(Kn) $$ 
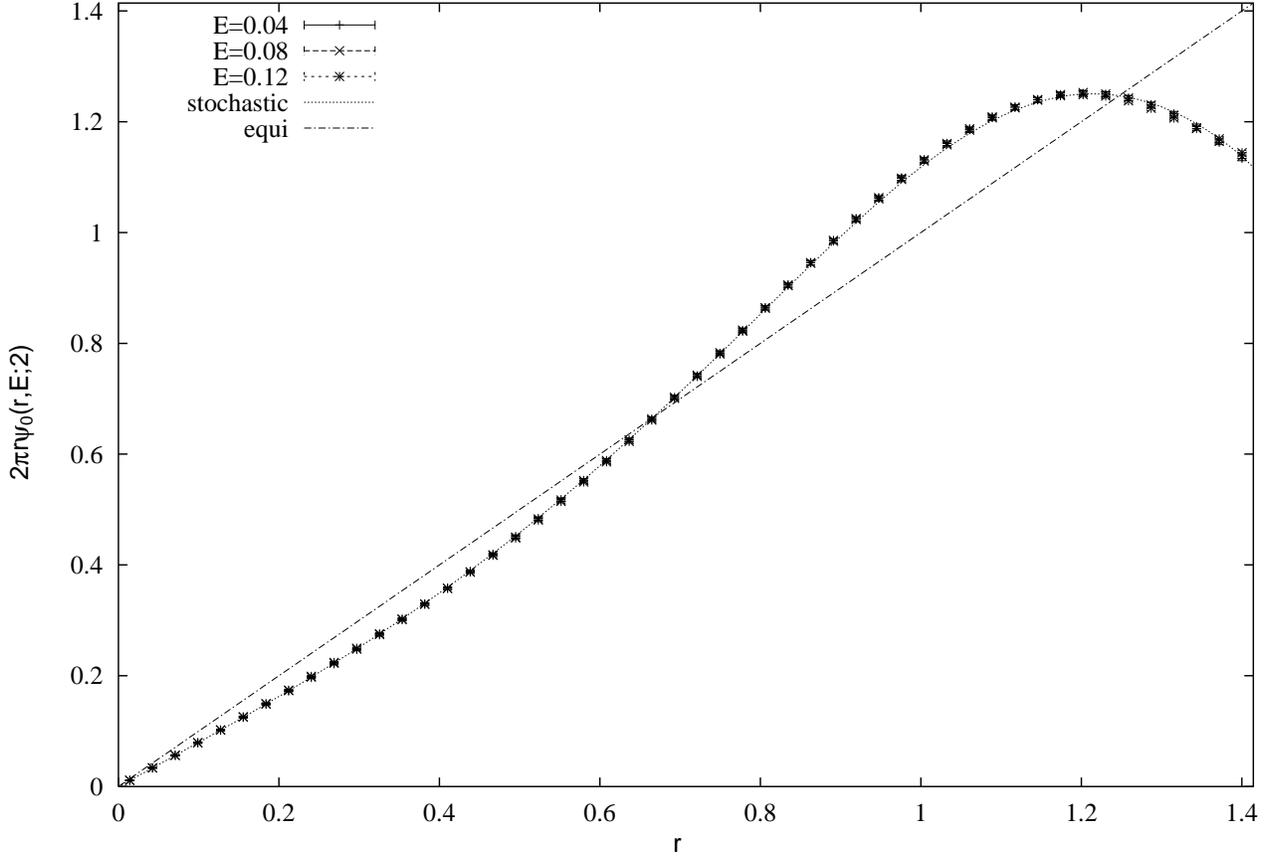
\gnuins fig5 fig5 {Plot of $2\pi r\psi_0(r,E;2)$  for 
different values of $E$. The straight dashed line is obtained from the
microcanonical distribution, Eq.\equ(mc). The dotted line gives the
result for the stochastic model}
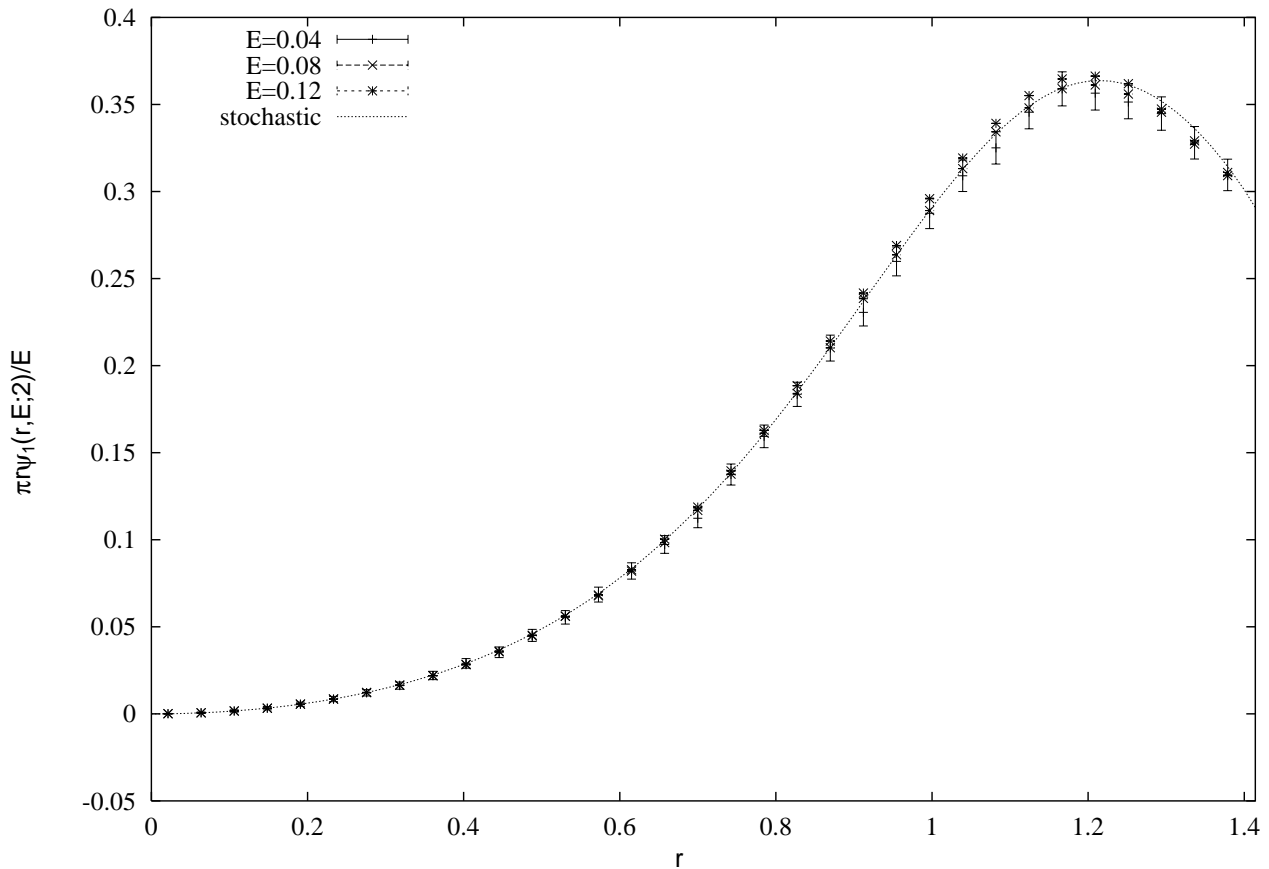
\gnuins fig6 fig6 {Plot of $\pi r\psi_1(r,E;2)/E$
for different values of $E$. The dotted line gives the result for the
stochastic model}

In Fig. \equ(fig5) we plot {\bf $2\pi r\psi_0(r,E;2)$}
for $E=0.04,0.08,0.12$ while Fig. \equ(fig6) is a plot of
$\pi r\psi_1(r,E;2)/E$ for the same values of the field. Both appear to
be almost independent of $E$ for those values of $E$ so we believe that
Figs. \equ(fig5) and \equ(fig6) represent a good approximation for the
limiting behavior $E\to 0$. Observe that, due to the symmetry $E\to
-E$ we expect the corrections to these functions to be of $O(E^2)$.
For comparison we also plotted there the results obtained analytically
from the stochastic model discussed in the Introduction and in section
3.

In Fig. \equ(fig5) we also plot the
`` microcanonical'' density of $|\bv_1|$ obtained from the
microcanonical ensemble of 2 particles with $\bv_1^2+\bv_2^2=2$. 
The microcanonical one particle density $f_{\rm micro}(\bv)$
is of course isotropic and the speed distribution,
 $2\pi|\bv_1|f_m (|\bv_1|,E=0;2)$, is

$$2\pi|\bv_1| f_m(|\bv_1|,E=0;2)={1\over \pi}|\bv_1| 
\int\delta(\bv_1^2+\bv_2^2-2)d\bv_2= |\bv_1|\ {\cal H}(2 - \bv_1^2) \
,\Eq(mc)$$
where ${\cal H}(x)$ is the Heaviside function.  This is seen to be
very different from what we obtain from our simulations or
analytically from the stochastic model for $E\to 0$.  We did a similar
analysis for $N>2$ and in Figs. \equ(fig7) and
\equ(fig8) we present the corresponding results for $N=50$.

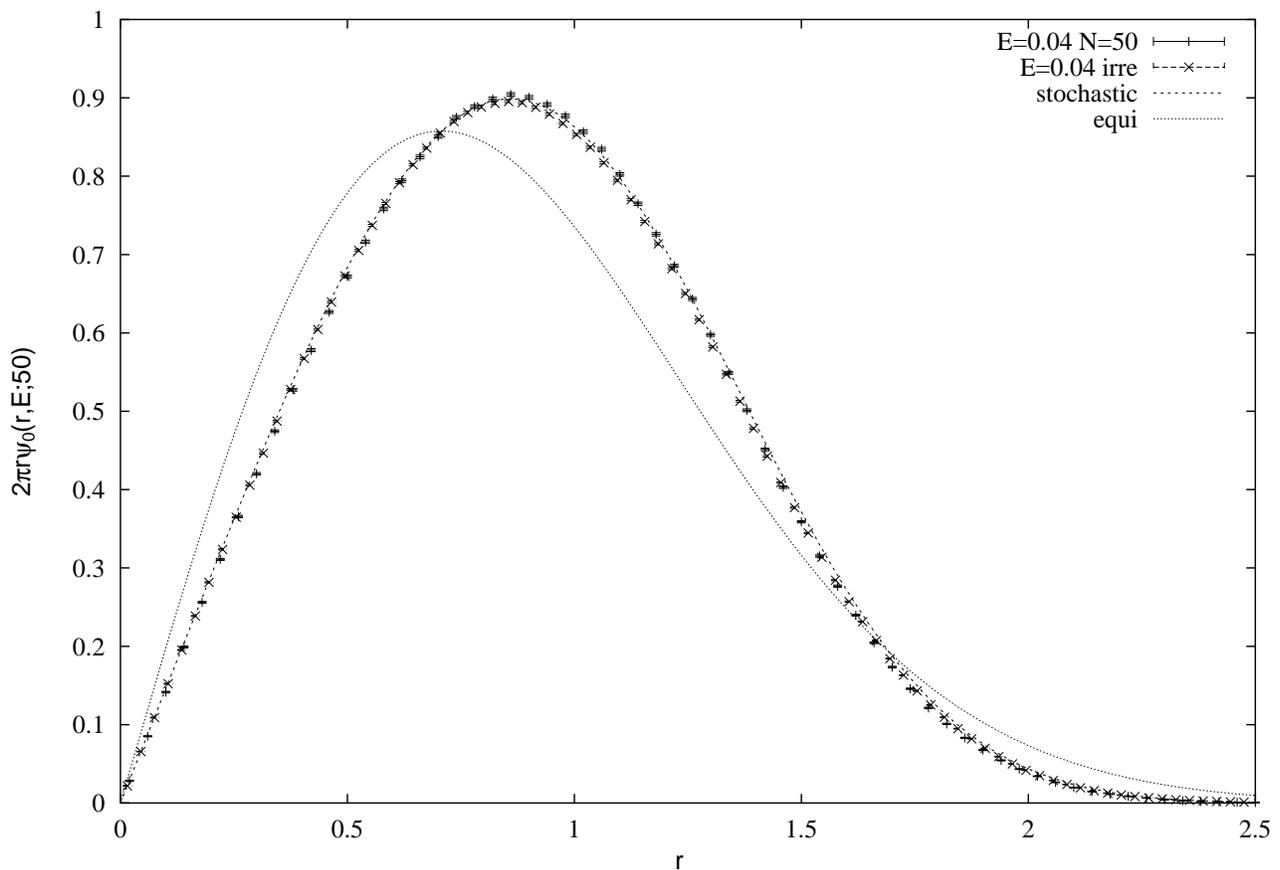
\gnuins fig7 fig7 {Plot of $2\pi r\psi_0(r,E;50)$ 
for $E=0.04$. Also shown are the results from simulations of \equ(vi)
and from analytic solutions of the corresponding stochastic equation,
Eq.\equ(sol2). For comparison we also show the microcanonical result,
corresponding to a Maxwellian.}

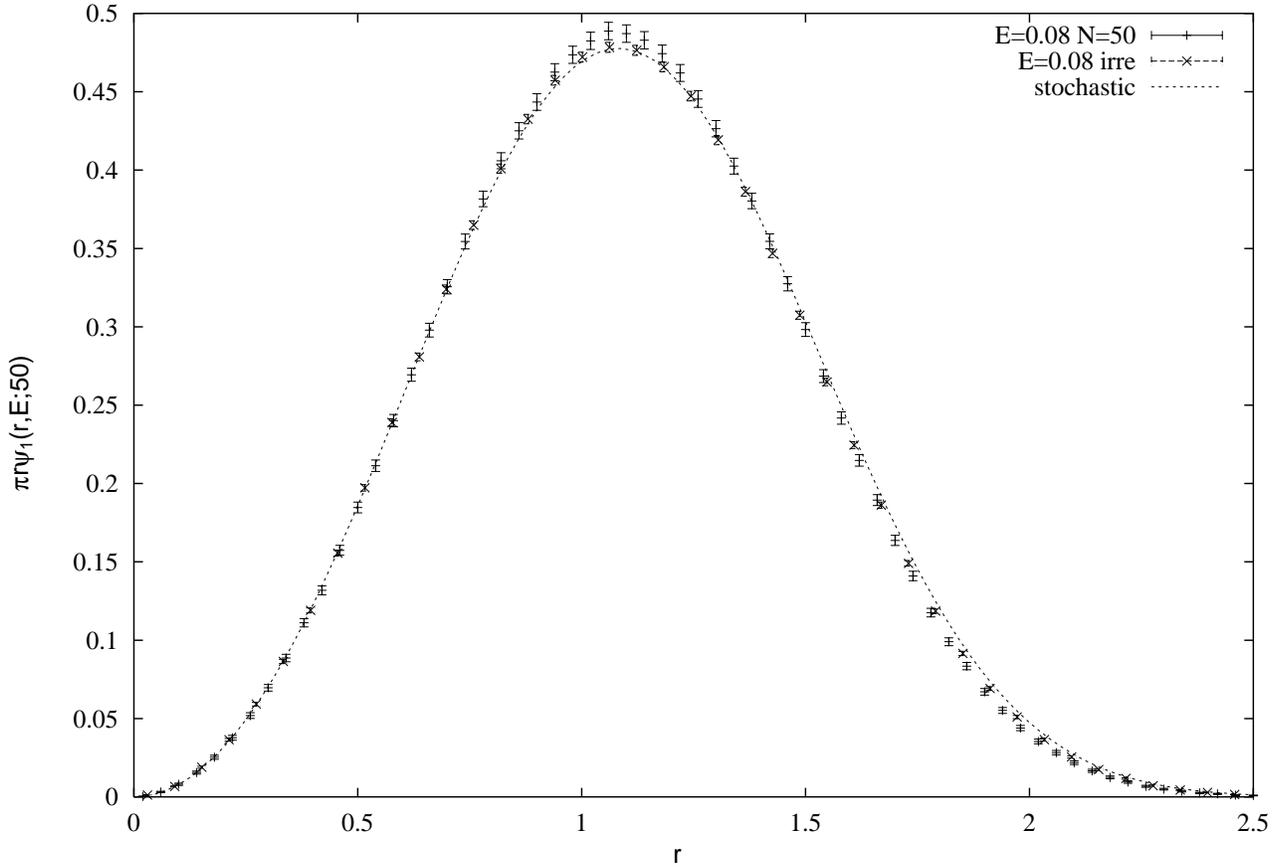
\gnuins fig8 fig8 {Plot of $\pi r\psi_1(r,E;50)/ E$ and comparison
with stochastic irreversible dynamics for $E=0.08$} 

\newsubsect{The $N=\infty$ limit}

As discussed in sec. 2.1, $\k(E,N)\to\tilde\k(E)$ as $N\to\i$. We
compared the $\tilde\k(E)$ obtained from our simulation, see
Fig. \equ(fig3), with that obtained from the irreversible
eq.\equ(vi). A way to do this self-consistently would be to choose the
parameter $\lambda$ in eq.\equ(vi) such that

$$\hat U(E)=\int d\bv |\bv|^2 \hat f^+(\bv,E)=1$$
and show that for this value of $\lambda$ the conductivity $\hat
\k(E)$ for the system described by eq.\equ(vi) is equal to $\tilde
\k(E)$.  Rather than doing this, we took the $\tilde\k(E)$ deduced
from the simulations as in Fig.\equ(fig3) and used it to determine
$\lambda$, i.e. we set $\lambda=\tilde \k(E)E^2$ in eq.\equ(vi). We
then computed, via simulation of eq.\equ(vi), a new conductivity $\hat
\k(E)$.  In Fig.\equ(fig_fit) we compare $\hat \k(E)$ and
$\tilde\k(E)$. The agreement is very good. We observe that it follows
from eq.\equ(vi)that $E^2 \hat\k(E)/\hat U(E)=\lambda$ so that this
agreement also confirms the self-consistency discussed above.

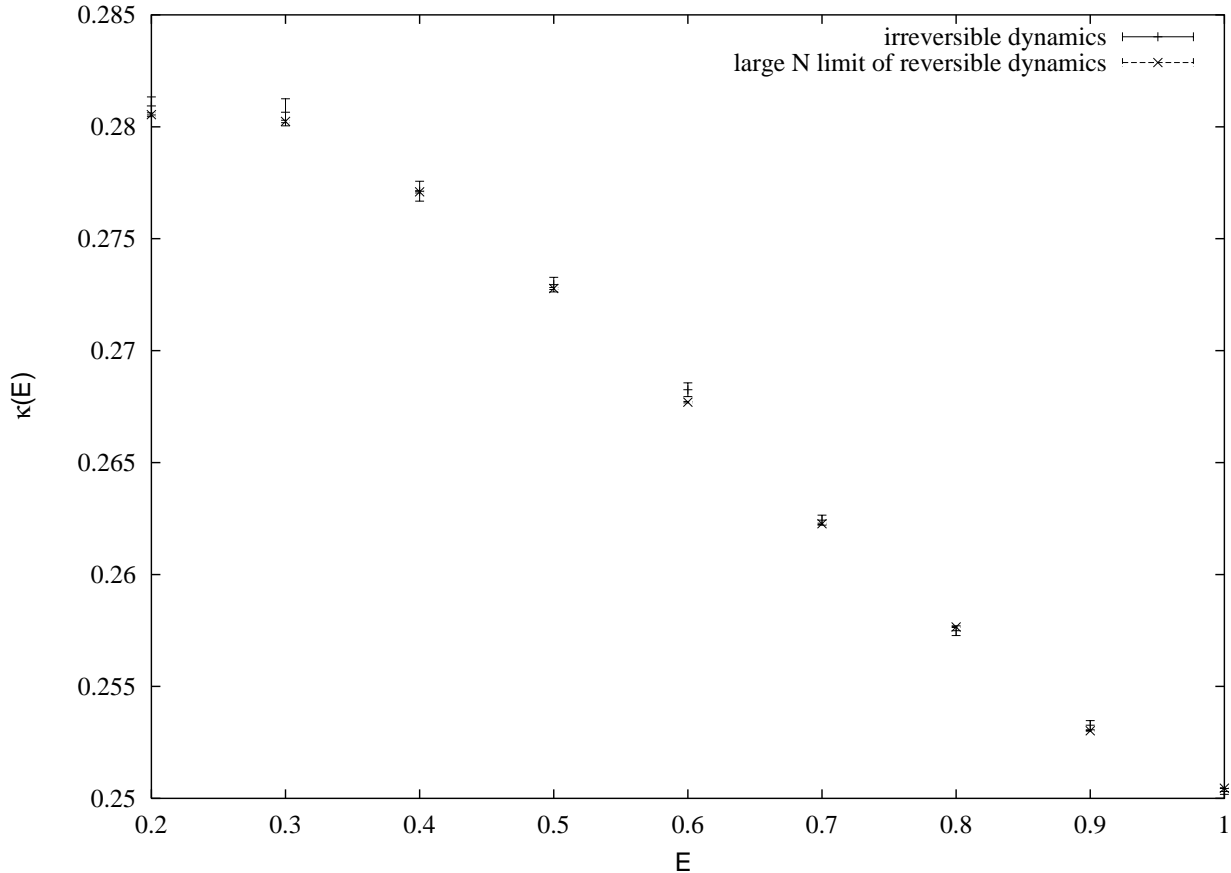
\gnuins fig_fit fig_fit {Comparison between the limiting value of the
conductivity $\tilde \k(E)$ in the reversible model and in the
irreversible model $\k_\infty(E)$.}

As for the reversible dynamics we can write 

$$\hat f^+(\bv,E)= \sum_{k=0}^{\infty}\phi_k(r,E) \cos k\theta\Eq(espi)$$

In Figs. \equ(fig7) and \equ(fig8) we compare $2\pi r\psi_0(r,E;50)$ and
$\pi r \psi_1(r,E;50)$ with $2\pi r\phi_0(r,E)$ and $\pi r \phi_1(r,E)$
respectively. The agreement is very good.  As we did for $N=2$ in
Fig. \equ(fig5) and Fig. \equ(fig6) we also plotted in
Figs. \equ(fig7) and \equ(fig8) the results obtained analytically from
the stochastic model discussed in the Introduction and in section
3. In Fig. \equ(fig7) we also plot the microcanonical density, \ie a
Maxwellian with $<\bv_1^2>=1$.

\newsec{Thermostatted Stochastic Evolution}

We now describe more precisely the stochastic model system in which
the collisions between particles and obstacles are replaced by
independent random scattering events. The model is specified by
writing the right hand side of eq. \equ(bltz), the evolution equation
for the $N$-particle phase space density of our system, which we now
call $F_t(\bQ,\bV)$, to distinguish it from the mechanical
$\rho_t(\bQ,\bV)$, as

{\bf $$\left({\partial F(\bQ,\bV,\bE)\over \partial t}\right)_{\rm coll}=
l^{-1}\sum_{i=1}^N\int_{(\bn\cdot \bv_i)<0} {(\bv'_i\cdot \bn)\over 2}
\left(F(\bQ,\bV_i',\bE)-F(\bQ,\bV,\bE)\right)d\bn\Eq(coll)$$} 
In \equ(coll) $\bn$ is a unit vector in the direction of the momentum
transfer in a ``collision'', $\vert \bn\vert=1$,
$\bv'=\bv-2\bn(\bn\cdot\bv_i)$ and $\bV_i'$ is identical to $\bV_i$
except for its $i$-th component which is replaced by $\bv'_i$. The
coefficient $l^{-1}$ multiplying the collision term is the inverse of
the mean free path between collisions, a parameter to be specified.

Eq. \equ(bltz) together with \equ(coll) describes 
a Markov process in which particles change the
directions of their velocities as if they were undergoing independent
random collisions with ``phantom obstacles'' at a rate equal to
$l^{-1}|\bv|$ with a uniformly distributed impact parameter \cita{IP}.
Between collisions the particles move according to eq.\equ(dyn1).
This model can be thought of as, and presumably even proven to be, the
Boltzmann-Grad limit of our system: \ie, we place discs of radius $R$
randomly in a square of side $L$ with density $\rho$ and then take
$R\to 0$, $\rho\to\i$ such that $l={1\over 2\rho R}$ stays
constant, see \cita{B-G}.

This system will, like our mechanical system, eq.\equ(dyn1), conserve
energy, so setting $\sum\bv_i^2=N$ the evolution takes place on
$\SS_N$. By general arguments \cita{GLP}, \cita{GLP2} we expect that
this system will, for $E\not=0$ approach, as $t\to\infty$, a unique
stationary density $F(\bV,\bE;N)$ which will satisfy the equation

$$\sum_{i=1}^{N}{\partial\over \partial
\bv_i}\left\{\left[\bE-\bE\cdot\bJ\bv_i\right] F(\bV,\bE;N)\right\}=
\left({\partial F(\bV,\bE;N)\over \partial t}\right)_{{\rm coll}} \Eq(blN)$$

For small $E$ we expand $F(\bV,\bE;N)$ as  a formal power series in $\bE$:
{\bf $$F(\bV,\bE;N)=F(\bR,\Theta)=\sum_{n=0}^{\infty}E^n F^{(n)}(\bR,
\bTheta) \Eq(sam)$$} 
where we have set $\bv_i=(r_i \cos \theta_i,r_i \sin \theta_i)$ and
$\bR =(r_1,\ldots, r_N)$, $\sum_i r_i^2=N$, $\bTheta =(\theta_1,
\ldots,\theta_N)$. Observe that in this way we get a singular perturbation
problem because $E$ multiplies the highest order derivative in
eq.\equ(blN).  Moreover $F^{+}(\bV,\bE;N)$ clearly depends only on $E/l$
so that we can, for the time being, set $l=1$.  Finally we can write,
as in the previous section,

$$F^{(n)}(\bR,\bTheta)=\sum_{\bk\in \zzz^N_+}
F^{(n)}(\bR,\bk)\prod_{i=1}^N\cos(k_i\theta_i)
\Eq(fou1)$$
where we have again used the symmetries of the problem.

Substituting \equ(fou1) into \equ(sam) one gets a hierarchy of
equations linking $F^{(n)}(\bR,\bk)$ to $F^{(n-1)}(\bR,\bk^i)$ where
$\bk^i=(k_1,\ldots,k_i+1,\ldots,k_N)$. From this, and from the fact
that the kernel of the collision operator depends only on $\bR$ we get
that $F^{(n)}(\bR,\bk)=0$ if $|\bk|>n$.  $F^{(0)}(\bR, 0)$ satisfies
the relation:

$${\partial\over \partial {r_i}}F^{(0)}(\bR,0)={4\over 3}{r_i}
F^{(1)}(\bR,0^i)\Eq(1or)$$ 
while for $F^{(1)}(\bR,0^i)$ we get the equation

$$\sum_i\left\{\left(-{r_i\over U}-{1\over
r_i}\right)F^{(1)}(\bR,0^i)+{\partial\over
\partial {r_i}}F^{(1)}(\bR,0^i)\right\}=0 \Eq(2or)$$
with $U=\sum_i r_i^2$. Equations \equ(1or) and \equ(2or) are easily
solved and, together with the fact that $F^{(1)}(\bR,0)\equiv 0$ give
us $F(\bR,\Theta)$ to first order in $E$

$$F(\bR, \bTheta)=C\delta(\sum_{i=1}^Nr_i^2-N)\left[{1\over \left(\sum_i
r_i^3\right)^{{2N-1\over 3}}}+{3(2N-1)E\over 4}
{r_i\cos\theta_i\over \left(\sum_i
r_i^3\right)^{{2N+2\over 3}}} +O(E^2)\right]\Eq(sol1)$$
where C is a normalization constant.  It is possible to write out the
full hierarchy of equations for $F^{(n)}(\bR,\bk)$ and see that they
can be solved iteratively but it is not clear that this is useful. We
shall therefore use eq.\equ(sol1) to compare with our numerical data
for small values of $E$. To do so we define the one particle
distribution $\tilde{f}(\bv,E;N)$ and develop it in a Fourier series
exactly as in eq.\equ(esp):

$$\tilde f(\bv,E;N)=\int d\bv_2\cdots d\bv_N\tilde
F(\bV,E;N)=\sum_{k=0}^\i \tilde \psi_k(r,E;N)cos (k_i\theta_i)\Eq(proj)$$

Before doing any comparisons we consider the stochastic version of the
$f_i(\bv,E)$ obtained from the irreversible dynamics defined by
eq.\equ(vi). Putting $\lambda=E^2\nu$, $\nu$
to be set to $\bar{\k}(E)$
when compared with the deterministic model, we get

$${\partial\over \partial
\bv}\left\{\left[\bE-E^2\nu\bv\right]\tilde f_i(\bv,\bE)\right\}=
\left({\partial \tilde f_i(\bv,\bE)\over \partial t}\right)_{{\rm
coll}} \Eq(bltz2)$$
where the collision term is again given by eq.\equ(coll) with $N=1$. 
Observe that
although eq.\equ(bltz2) contains three parameter ($E$, $\nu$ and $l$)
it depends only on $El$ and $\nu l^{-1}$.  Developing $\tilde
f_i(\bv,\bE)$ in a power series in $E$ we obtain in analogy to
\equ(sol1)

$$\tilde{f}_i(\bv,\bE)= C e^{-\frac{8}{9l}\nu r^3} (1+ 2\nu E r
\cos\theta)+O(E^2)\Eq(sol2)$$
where $C$ is a normalization constant. 

To compare $\tilde{f}_i (\bv,\bE)$ with the large $N$ limit of
$\tilde{f}(\bv,\bE;N)$ given in \equ(proj) and \equ(bltz2) we need to
fix the parameter $\nu$ (setting $l=1$).  This can be done
self-consistently requiring that:

$$\int|\bv|^2\tilde f_i(\bv,\bE)d\bv=1\Eq(self)$$
Solving eq.\equ(self) for $\nu$ and using it to compute $\tilde f_i$
we expect that:

$$\lim_{N\to\i}\tilde  f(\bv,E;N)=\tilde f_{i}(\bv,E)\Eq(equi)$$
While we have not proven this equivalence we believe that it should
follow from general considerations: it would follow formally from
showing that, in the limit $N\to\i$, $\tilde{F}(\bv,\bE;N)$
factorizes, as is usually the case for systems with mean field type
interactions. This is certainly consistent with our numerical results.

\newsec{Comparison between the deterministic and stochastic time evolution}
\newsubsect{The distribution of the modulus of $\bv$}

For $N=1$ the exact solution, for $E=0$, of both the stochastic and
mechanical models is $f(\bv,0;1)=\delta(\bv^2-1)$.  For $N=2$, we are able
to compute the one particle distribution from eq.\equ(sol1). This
yields

$$r\tilde\psi_0(r,E;2)={Cr\over r^3+(2-r^2)^{3/2}}+O(E^2)\Eq(dist2)$$
where $C$ is a normalization constant. This is plotted in {\bf
F}ig. \equ(fig5) and one can easily see that the agreement with the
numerical solution of the deterministic model is very good.

A similar agreement is obtained for $N=5$ although, as already said we
were not able to integrate eq.\equ(proj) for $N>2$ so that we computed this
integral numerically by simulating the process associated to eq.\equ(bltz) 
with collision term given by eq.\equ(coll).

Finally for $N=50$ we see in  Fig. \equ(fig7) that our deterministic
\equ(dyn1), stochastic \equ(bltz2) and irreversible \equ(vi) models 
give indistinguishable results. This certainly suggests the
validity of \equ(vi) and \equ(equi) for large $N$.

\newsubsect{The first Fourier component of the distribution of $\bv$}

The analysis of the first Fourier component of the distribution of
$\bv$ is less straightforward because we must fit the parameter $l$
appearing in eq.\equ(coll). In the stochastic system $l$ represent the
mean free flight of a particle. The concept of mean free flight is not
uniquely defined for the mechanical model.  For this reason we used
$l$ as a fitting parameter for matching $\tilde{\psi}_1(r,E;N)$ with
$\psi_1(r,E;N)$. We will go back to the mechanical meaning of this
parameter in the following section. The case $N=2$ is reported in
Fig.\equ(fig6) where, for the periodic case, we used a field $E=0.04$
and for the stochastic one we have the expression

$$r\tilde\psi_1(r,E;2)=
{1\over 2}{9El\over 4}{Cr^2\over \left(r^3+(2-r^2)^{3/2}\right)^2}+O(E^3)
\Eq(dist3)$$
with $C$ the same costant appearing in eq.\equ(dist2) The agreement is
again very good and we obtain from the fit $l=0.46$ (in the unit
discussed in the introduction). As in the previous case we did the
same comparison for 5 particles, obtaining again a very good
agreement. Moreover also in this case the value of $l$ is very close
to that obtained for $N=2$.  Finally it is interesting to check if
this agreement remains when $N\to\infty$, \ie for the stochastic
irreversible equation \equ(bltz2). As can be seen from Fig.\equ(fig8)
the agreement is again very good and we still get the same value for
the parameter $l\simeq 0.46$.

We were also able to compute $\psi_k(r,E;2)$ and $\phi_k(r,E)$ for
$k=2$ and $3$. It is also easy to compute the lowest order
contribution to $\tilde\psi_k(r,E;2)$ and $\tilde\phi_k(r,E)$,
extending the computation from section 3. It is thus possible to
compare, at least in this limited situation, the results.  Contrary to
what we found for $k=0$ and 1, $\psi_2(r,E;2)$ is quite different from
$\tilde\psi_2(r,E;2)$. Analogously $\phi_2(r,E)$ and
$\tilde\phi_2(r,E)$ differ significantly. A comparison of the term
with $k=3$ also shows deviations between the mechanical and the
stochastic models although, surprisingly, much smaller than those
found for $k=2$. We note however that for this comparison we only have
data for $E=0.012$.

\newsubsect{The mean free flight.}

In kinetic theory one can define the mean free flight in two ways. Denoting by
$\ell_i(X)$ the distance travelled by particle $i$ before its first
collision with an obstacle starting form the point $X\in\SS_N$,
$l_0$ is the average of $\ell_i(X)$ with respect to the SRB
distribution $\mu^+(dX,\bE;N)$ (it clearly does not depend on $i$). On
the other hand we can consider the set $\SS^i_N$ of points such
that particle $i$ is undergoing a collision, \ie $\bq_i$ is on the
boundary of one of the scatterers, then $l_1$ is the average of
$\ell_i(X)$ on $\SS^i_N$ with respect to the projection of the SRB
distribution $\mu^+(dX,\bE;N)$. Observe that for the stochastic model
these two quantities are identical.  

We computed both $l_0$ and $l_1$ for the mechanical system with
$N=2,5,50$ and for the irreversible dynamics eq.\equ(vi) with
$E=0.04$. This was done by running a very long trajectory and taking
the average of the distance travelled by a particle between two
collisions to compute $l_1$ or numerically integrating $\ell_i(X)$
along the trajectory to compute $l_0$. The results appears to be
independent of $N$, at least within the accuracy of our
computations, and are:

$$\eqalign{l_0&=0.46\cr l_1&=0.58}$$

The value of $l_0$ agrees very well with the value obtained from the
fit of $l$ reported in the previous section. This implies that the
correct way to compare the stochastic and the mechanical model is to
use $l_0$ as the mean free flight parameter in eq.\equ(coll).
This is consistent with the Green-Kubo formula eq.\equ(GK). We
saw in sect. 2.1 that eq.\equ(GK) is well verified for the
conductivity at small field of the deterministic model. In the case of
the stochastic model eq.\equ(GK) reduces to an integral relation
between $F^{(0)}(\bR,0)$ and $F^{(0)}(\bR,0^i)$, see
eq.\equ(1or)\equ(2or) in sect. 3. We did not prove this identity although
numerical analysis for small $N$ seems to verify it. Finally the
agreement between $\psi_0(r,0;N)$ and $\tilde\psi_0(r,0;N)$ observed
in sect.4.1 tells us that the ratio between the conductivity for the
deterministic and stochastic dynamics is independent of $N$ at least
for $E\to 0$. From eq.\equ(sol1) we know that the conductivity for the
stochastic model with one particle and $E=0$ is $3l/4$ so that also
for the deterministic model we have

$$\k(0,1)={3\over 4}l_0\Eq(agr)$$
This relation is also very well verified by our computation for the 1
particle system. 

To better compare the deterministic and stochastic models we also
computed the distribution $P(\ell,\bE;N)$ of $\ell_i(X)$ with respect
to the SRB distribution.  This distribution for 5 particles and
$E=0.04$ is shown in Fig.\equ(fig13) together with an exponential law
with the same average, \ie the distribution one would obtain running
the same simulation for the stochastic case. We did similar
computation for $E=0.04$ and $N=2,10$ and $50$. The results are again
independent from $N$.

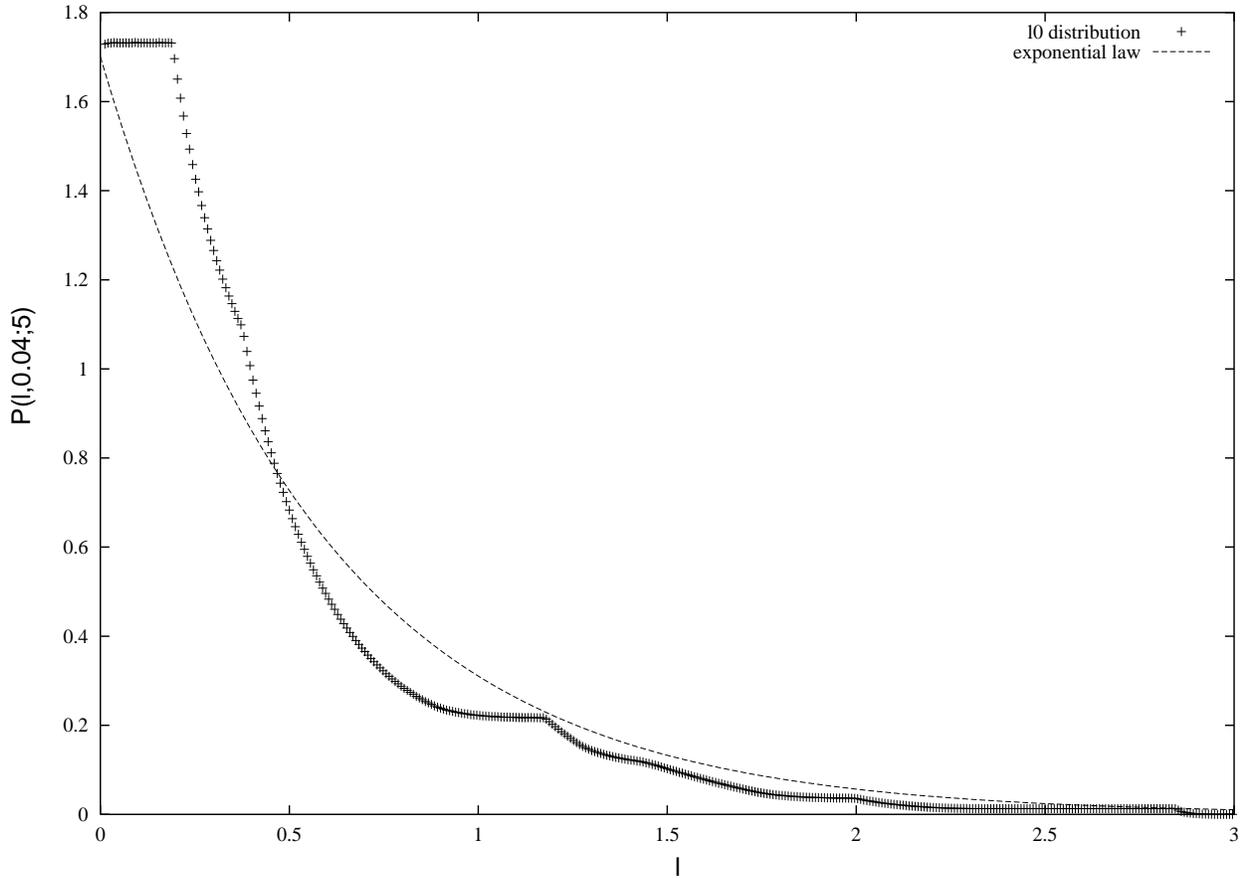
\gnuins fig13 fig13 {Free path distribution $P(l,0.04;5)$ compared with
an exponential distribution with the same average}

\newsec{Conclusions}

To put our study here in a physical context we note that a system of
noninteracting electrons moving under the influence of an external
electric field while undergoing elastic scatterings is often
used as a crude model of electrical conduction in metals (the Drude
model) \cita{AM},\cita{K},\cita{DR}. To obtain the conductivity the velocity
distribution function of the electrons is then computed from a
Boltzmann type equation like eq.\equ(blN): with $N=1$ and {\it
without} the thermostatting $\bE\cdot\bJ$ term. By doing this
calculation only to linear order in $E$ one avoids the problem that,
without the thermostat eq.\equ(blN) does not have a solution since the
system will never be in a true steady state \cita{PW}. A crucial
ingredient in the calculation is the explicit assumption that for
$E=0$ the distribution is one corresponding to equilibrium at a given
specified temperature $T$, \ie Maxwellian for a classical system. 

This description of the system of
independent electrons interacting with the lattice of ions only via
elastic collision is clearly not realistic. It is just used for
obtaining a simple quick answer for the zero (small) field
conductivity.
For a more complete description of the steady state in a conductor one
has to consider the system to be in contact with some {\it reservoir}
which will absorb the heat generated by the current. It is this
interaction with some external reservoir that was replaced, in the
model considered here, by an artificial thermostat. To our surprise
however we found that  this modeling does not lead to a Maxwellian
distribution when $E\to 0$ even when $N$ is very large. This means
that there is no {\it equivalence of ensembles} when it comes to
modeling how the energy is extracted from the system- at least when
there is no direct interactions between the particles other than that
induced by the thermostat. We expect (and have some indication
\cita{Ga}) that this will change when we include collisions between
the particles. Still it raises some caution about ``thermostats'' as a
model for the description of stationary nonequilibrium states.

\medskip
\0{\bf Acknowledgment.} We are indebted to G. Gallavotti, 
P.L. Garrido, S. Goldstein, A. Rohlenko,  D. Ruelle and particularly H. van Beijeren
for many helpful discussions and suggestions.
Much of the research was carried out at Rutgers University where it was supported in
part by NSF Grant DMR-9813268, and Air Force Grant F49620-98-1-0207. 
D. D. is Charg\'e de recherches at the FNRS.
V. R. was supported by the Foundation BLANCEFLOR Boncompagni-Ludovisi n\'ee Bildt.

\rife{AM}{AM}{N.W. Ashcroft, N.D. Mermin, {\it Solid States Physics}, 
Holt, Rinehart and Winston, New York (1976)}

\rife{BH}{BH}{W. Braun, K. Hepp, {\it The Vlasov dynamics and its fluctuations in the $\frac{1}{N}$ limit of interacting classical particles }, Commun. Math. Phys. {\bf 56}, 101-113 (1977)}

\rife{BDL}{BDL}{F. Bonetto, D. Daems, J.L. Lebowitz, {\it Properties of  
Stationary Nonequilibrium States in the Thermostatted Periodic Lorentz Gas I: the One Particle System},
Journ. Stat. Phys. {\bf 101}, 35-60 (2000)}

\rife{BGG}{BGG}{F. Bonetto, G. Gallavotti, P.L. Garrido, {\it Chaotic
principle: an experimental test}, Physica D {\bf 105}, 226--252 (1997).}

\rife{BKL}{BKL}{F. Bonetto, A.J. Kupiainen, J.L. Lebowitz,
{\it Perturbation theory for coupled Arnold cat maps: absolute
continuity of marginal distribution}, preprint.}

\rife{B-G}{G}{G. Gallavotti, {\it Rigorous theory of the Boltzmann equation in the Lorentz gas}, 
Nota Interna n.358, Istituto di Fisica, Universit\`a di 
Roma, 10 feb. 1972}

\rife{BS}{BS}{L. Bunimovich, Ya. Sinai, {\it Statistical properties of 
Lorentz gas with periodic configuration of scatterers}, Commun. Math. Phys. 
{\bf 78}, 479--497 (1980) }

\rife{IP}{C}{C. Cercignani, {\it The Boltzmann equation and its applications},
Springer-Verlag, New York (1988)}

\rife{CELS}{CELS}{N.I. Chernov, G.L. Eyink, J.L. Lebowitz,
Ya.G. Sinai, {\it Steady state electric conduction in the periodic Lorentz gas}, 
Commun. Math. Phys. {\bf 154}, 569--601 (1993).}

\rife{DM}{DM}{C.P. Dettmann, G.P. Morriss, 
{\it Proof of Lyapunov exponent pairing for system at constant kinetic
energy}, Phys. Rev. E {\bf 53}, 5545-5548 (1996).}

\rife{Ru}{ER}{J.M. Eckmann, D. Ruelle, {\it Ergodic Theory of Chaos and
Strange Attractors}, Rev. Mod. Phys. {\bf 57} 617--656 (1985).}

\rife{Ga}{G}{G. Garrido, Private communication}

\rife{GLP}{GLP}{S. Goldstein, J. Lebowitz, E. Presutti, {\it
Mechanical systems with stochastic boundaries}, Proceedings of
conference on random fields, 403--419,Esztergom, Hungary, June 1979,
Coll. Math. Soc. J\'anos Bolyai} 

\rife{GLP2}{GLP2}{S. Goldstein, J. Lebowitz, E. Presutti, {\it
Stationary Markov chains}, Proceedings of conference on random fields,
421--461,Esztergom, Hungary, June 1979, Coll. Math. Soc. J\'anos
Bolyai} 

\rife{K}{K}{C. Kittel: {\it Introduction to Solid States Physics}, Wiley, 
New York (1986)} 

\rife{Lan}{L}{O.E. Lanford, {\it Time dependents phenomena in
statistical mechanics}, Mathematical Problems in theoretical physics
(Proc. Internat. Conf. Math. Phys., Lausanne, 1979), 103-118} 

\rife{DR}{Lo}{H. Lorentz, {\it Collected Papers}, Martinus Nijhoff,
The Hague, 1936}

\rife{MH}{MH}{Moran Hoover, {\it Diffusion in a periodic Lorentz gas}, 
Journ. Stat. Phys. {\bf 48},  709--726 (1987).}

\rife{PW}{PW}{J. Piasecki, E. Wajnryb, {\it Long-Time Behavior of the
Lorentz Electron Gas in a Constant, Uniform Electric Field}, 
Journ. Stat. Phys. {\bf 21}, 549--559 (1979).} 

\rife{Ru2}{R1}{D. Ruelle, {\it A remark on the equivalence of isokinetic
and isoenergetic thermostats in the thermodynamic limit}, Journ. Stat. Phys. {\bf
100}, 757--763 (2000)}

\rife{Ru3}{R2}{D. Ruelle, {\it Smooth dynamics and new theoretical ideas
in nonequilibrium statistical mechanics}, Journ. Stat. Phys. {\bf 95}, 
393--468 (1999)}

\rife{S}{S}{H. Spohn, {\it Large scale dynamics of interacting
particles}, Spinger Verlag (1991)} 

\rife{VB}{vB}{H. van Beijeren, private communication}

\rife{VB1}{vB1}{H. van Beijeren, J. R. Dorfman, E. G. D. Cohen, 
H. A. Posch, and Ch. Dellago, {\it Lyapunov Exponents from Kinetic 
Theory for a Dilute, Field-Driven Lorentz Gas}, Phys. Rev. Lett. 
{\bf 77} 1974--1977 (1996)}

\rife{VL}{VL}{M Wojtkoski, C. Liverani, {\it Conformally symplectic
dynamics and symmetry of Lyapunov spectrum}, Commun. Math. Phys. (1997)
in print (Jan. 30).} 

\biblio
\fine{}
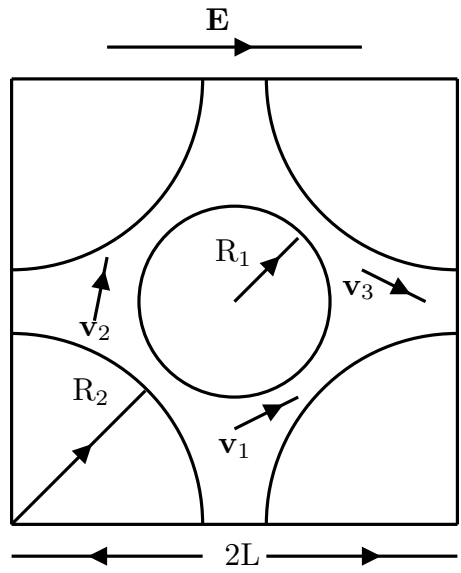
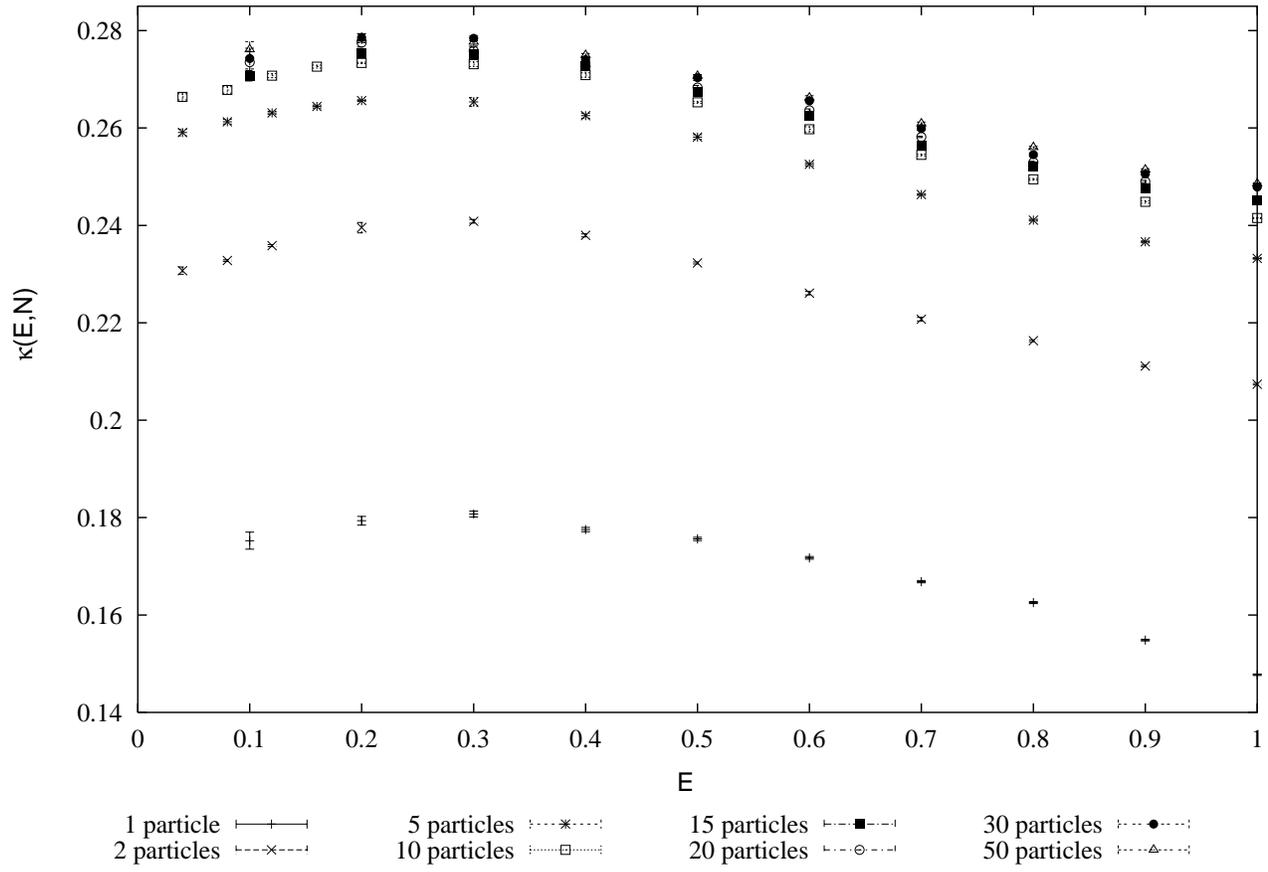
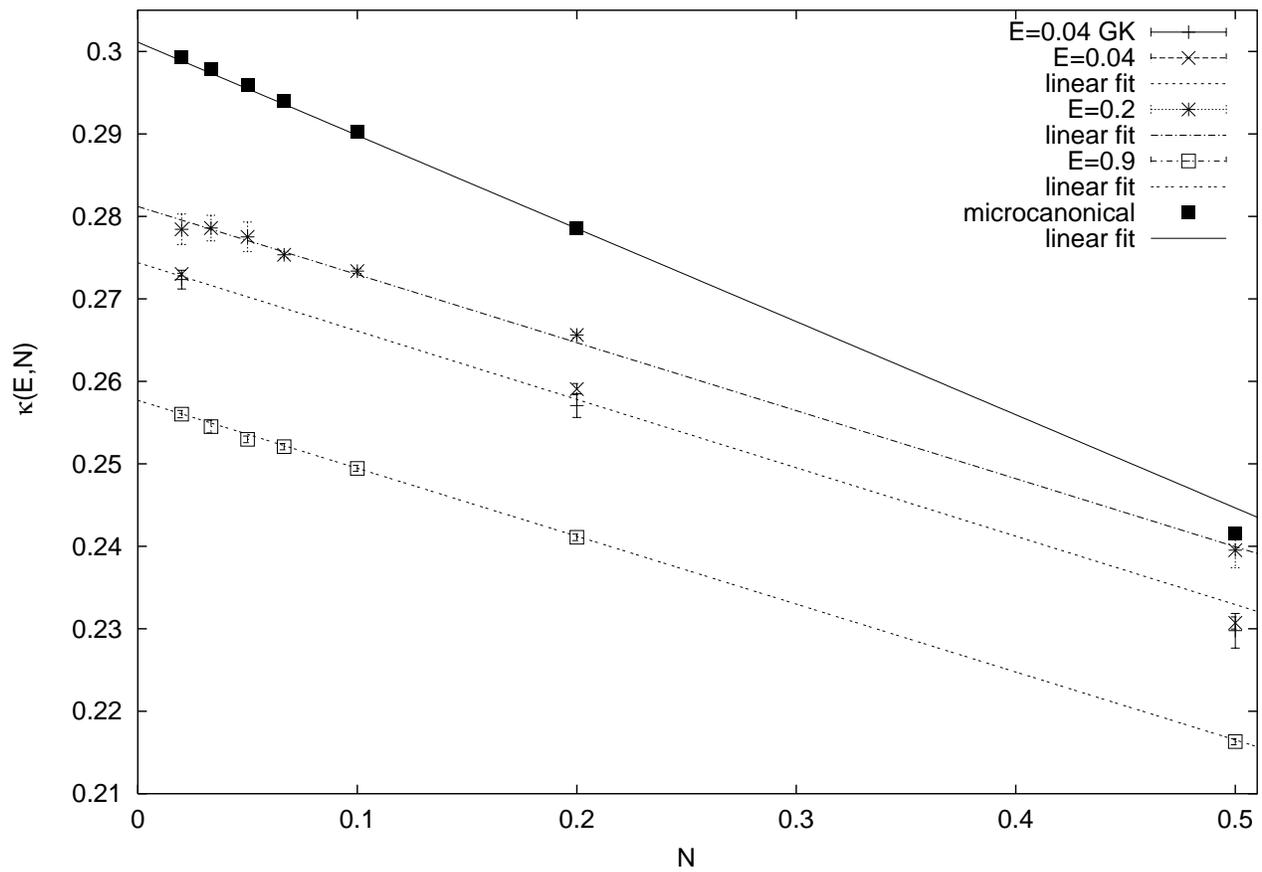
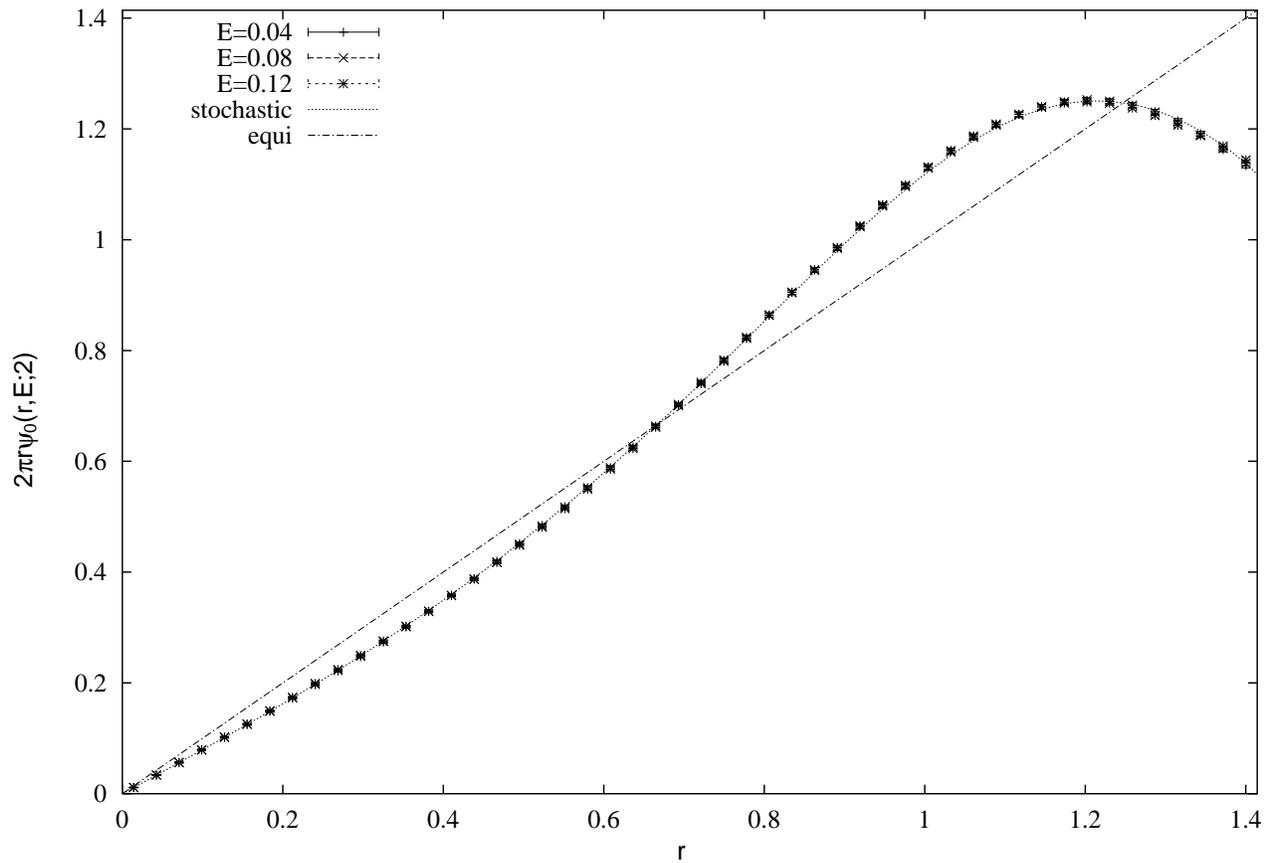
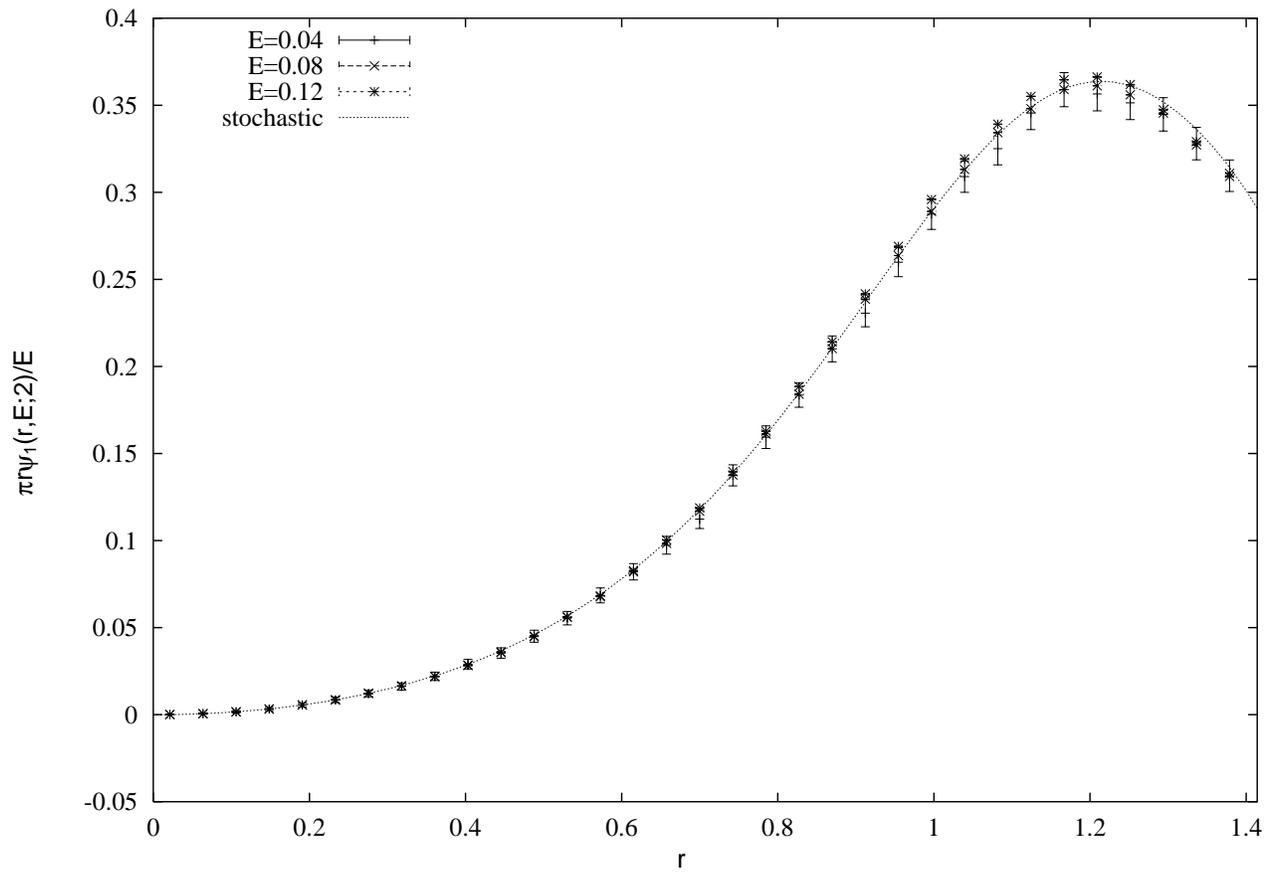
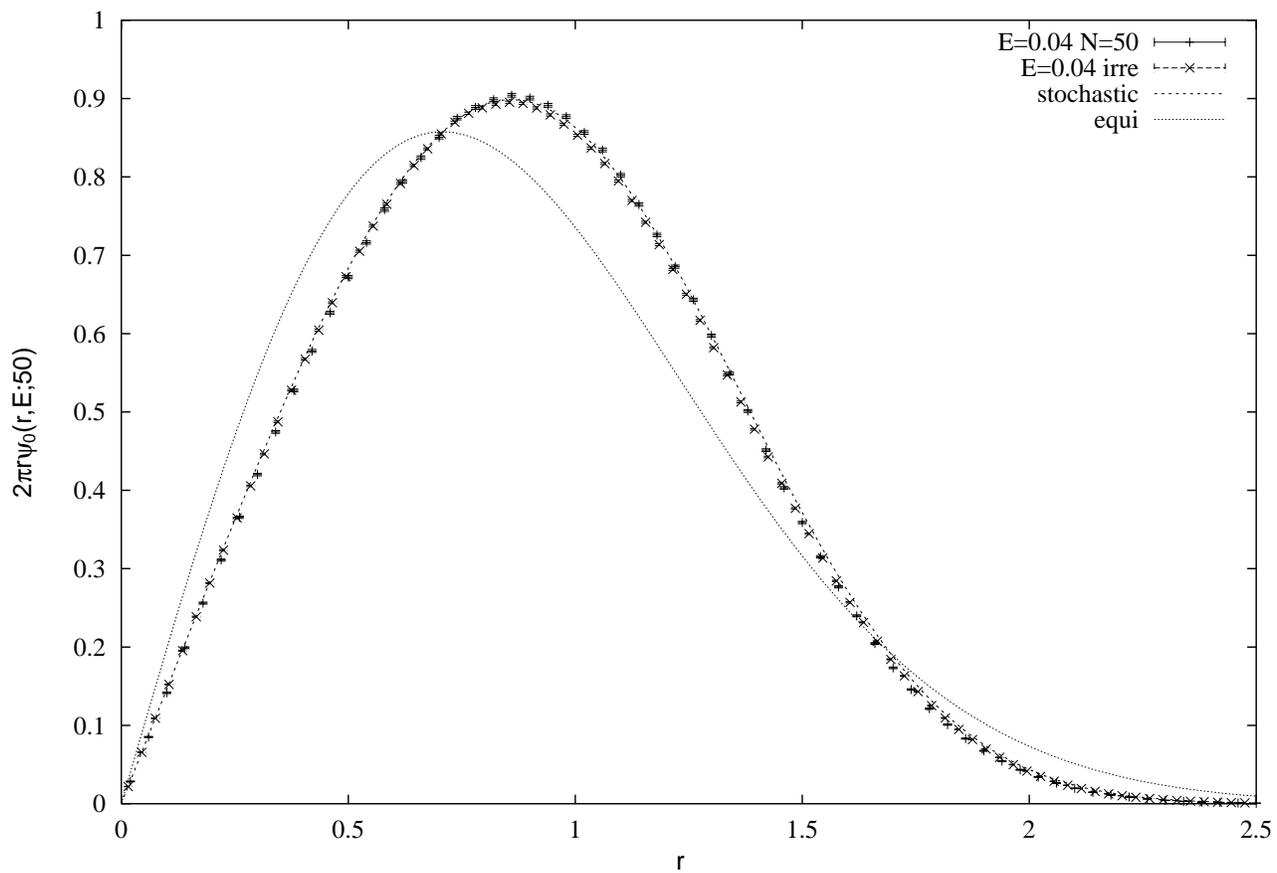
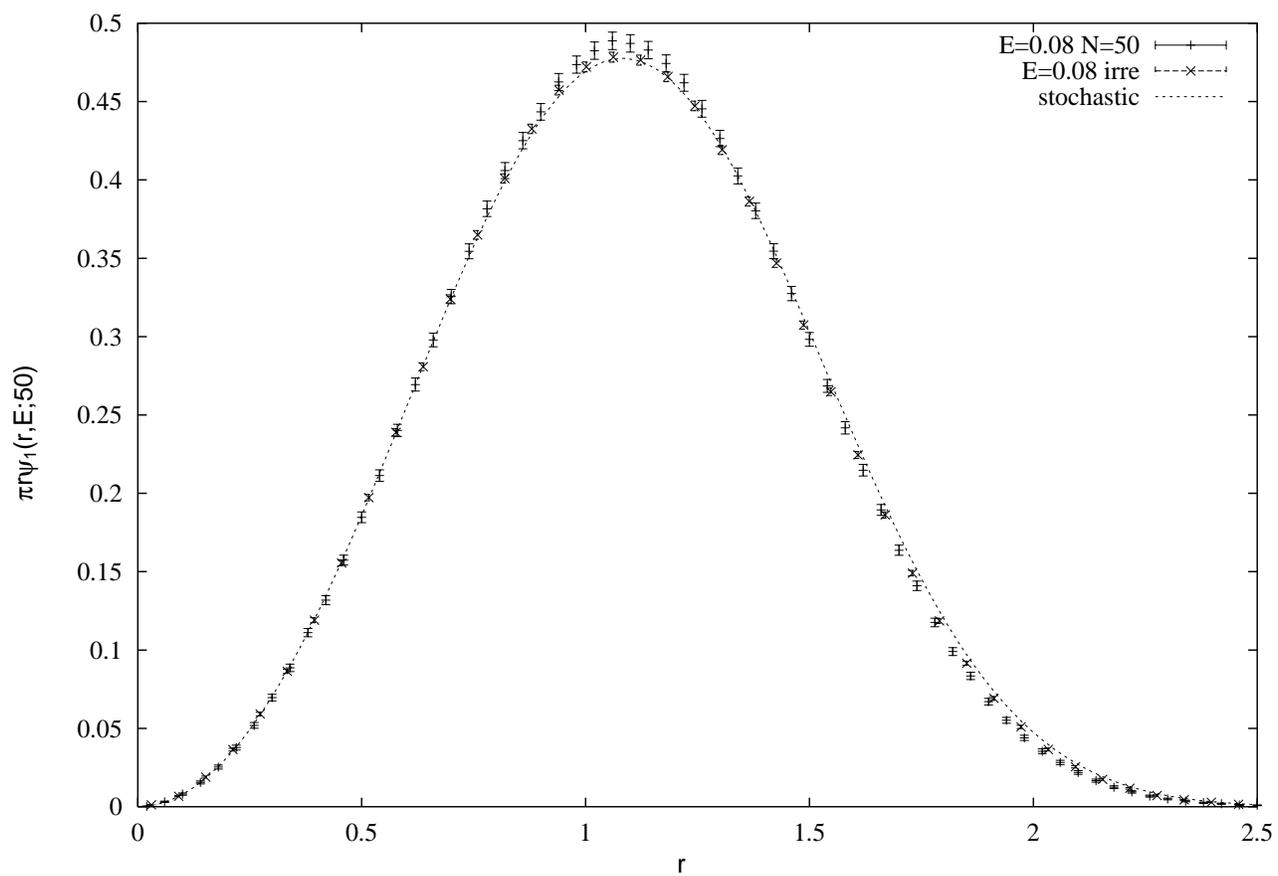
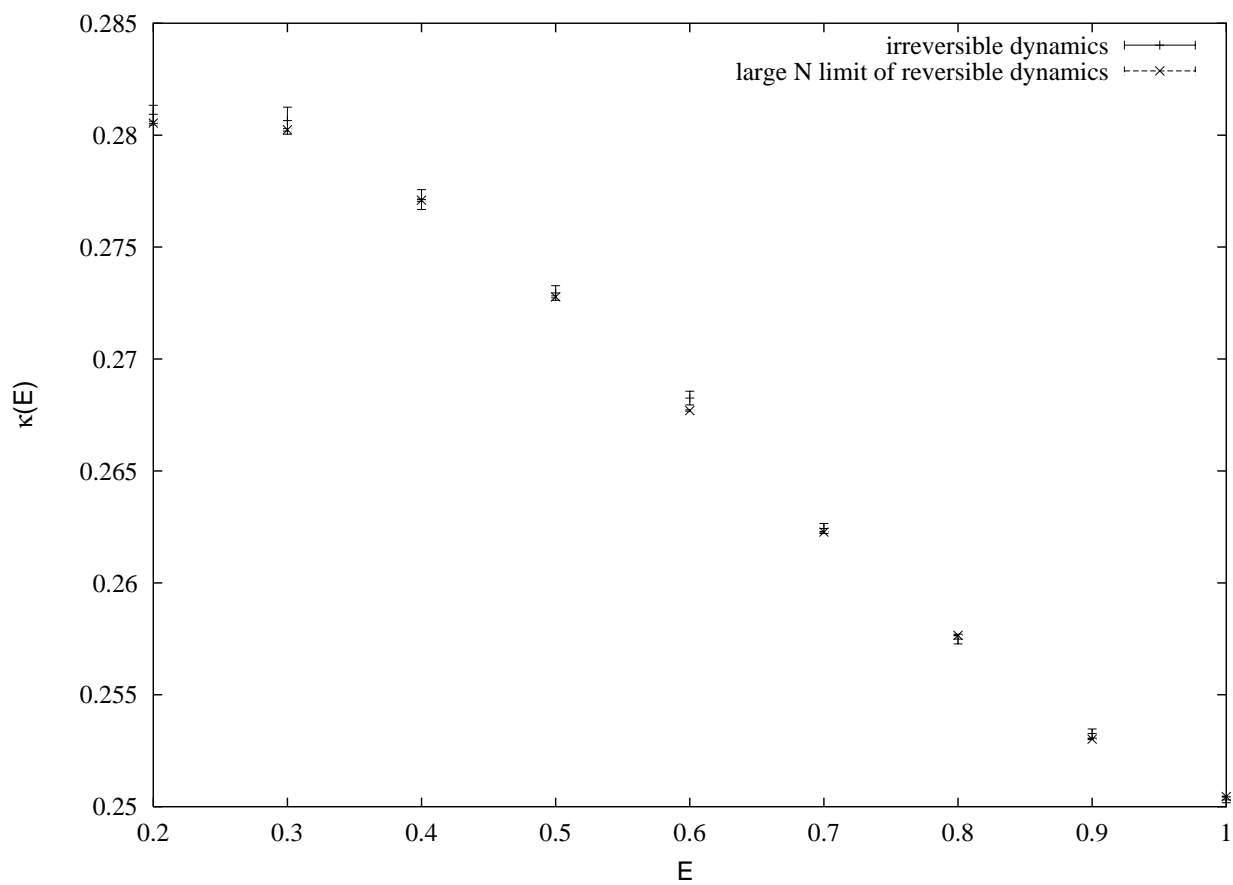
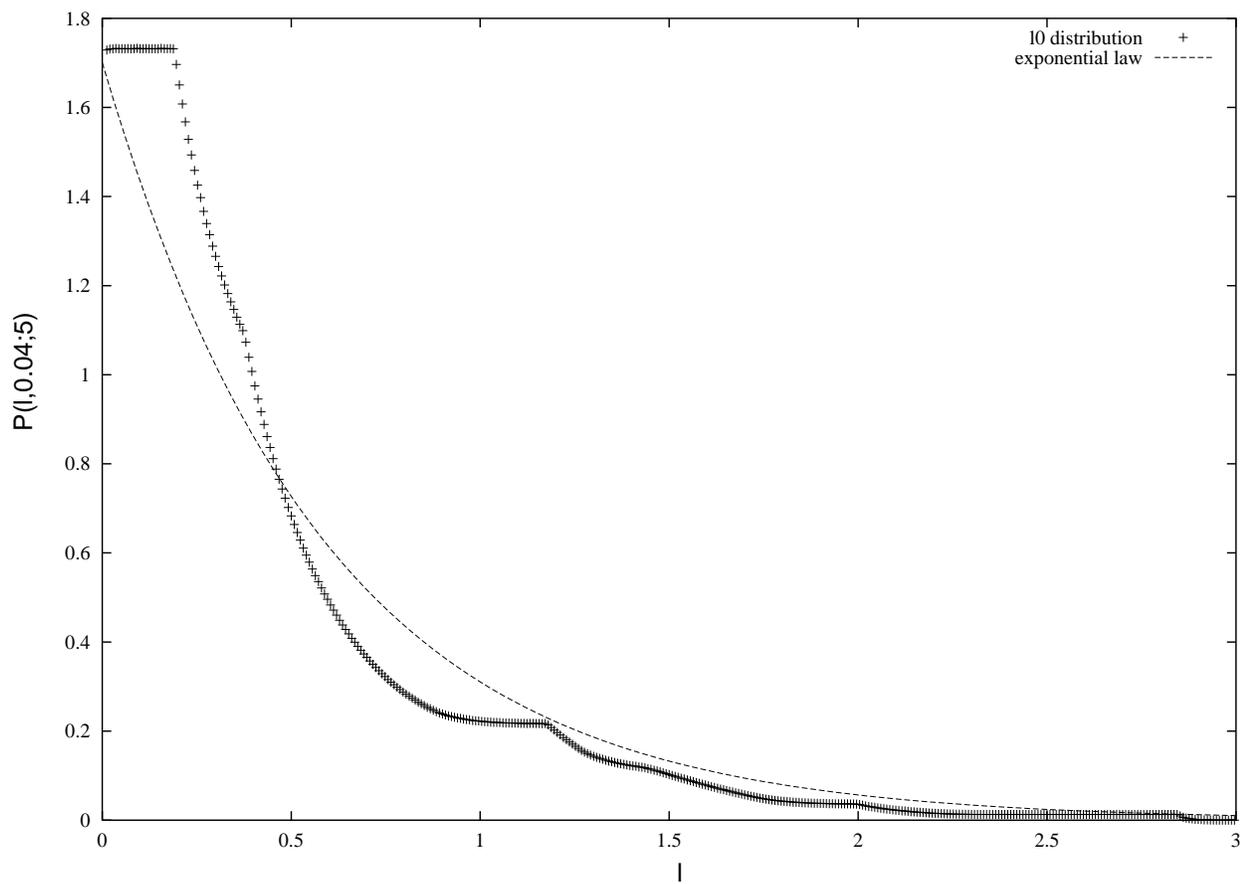
\end